# METABOLIC ALLOMETRIC SCALING OF MULTICELLULAR ORGANISMS AS A PRODUCT OF EVOLUTIONARY DEVELOPMENT AND OPTIMIZATION OF FOOD CHAINS

YURI K. SHESTOPALOFF

Production of energy is a foundation of life. Metabolic rate of organisms (amount of energy produced per unit time) generally increases slower than organisms' mass, which has important implications for life organization. This phenomenon, when considered across different taxa, is called interspecific allometric scaling. Its origin has puzzled scientists for many decades, and still is considered unknown. In this paper we posit that natural selection, as determined by evolutionary pressures, leads to distribution of resources, and accordingly energy, within a food chain, which is optimal from the perspective of stability of the food chain, when each species has sufficient amount of resources for continuous reproduction, but not too much to jeopardize existence of other species. Metabolic allometric scaling is then a quantitative representation of this optimal distribution. Taking locomotion and the primary mechanism for distribution of energy, we developed a biomechanical model to find energy expenditures, considering limb length, skeleton mass and speed. Using the interspecific allometric exponents for these three measures and substituting into the locomotion-derived model for energy expenditure, we calculated allometric exponents for mammals, reptiles, fish and birds, and compared these values with allometric exponents derived from experimental observations. The calculated allometric exponents were nearly identical to experimentally observed exponents for mammals, and very close for fish, reptiles and the basal metabolic rate of birds. The main result of the study is that the metabolic allometric scaling is a function of a mechanism of optimal energy distribution between the species of a food chain. This optimized sharing of common resources provides stability of a food chain for a given habitat and is guided by evolutionary pressures and natural selection.

*Keywords*: food chains; energy distribution; locomotion; speed; mammals; fish; reptiles; birds

## 1. Introduction

*Note*: The presented study, considering the origin and causes of metabolic allometric scaling, is described in two articles. This is the main article, which should be read first. It explores multicellular organisms. The second article considers origin of metabolic allometric scaling in unicellular organisms (Ref. 18).

### 1.1. *Problem's background*

Numerous biochemical processes, supporting life existence, its evolution and reproduction, rely on production of energy from acquired nutrients (meaning all kinds of involved substances, including mineral components). In many instances, electromagnetic radiation in a visible or nearby wavelength spectrum is also required, like sunlight for photosynthesis in plants. Most common energy producing biochemical mechanisms use oxygen, although there are many organisms, which employ anaerobic or both aerobic and anaerobic biochemical reactions for energy production. Due to their importance for biological, medical, biotechnological and other applications, energy producing mechanisms are intensively studied from different perspectives, at all scale levels - from molecules to whole organisms to systems of organisms in different strata.

One of the important directions of such studies is a metabolic allometric scaling (MAS), to which allometric studies of other organismal properties often relate, such as scaling of size of limbs, organs, morphological and kinematic features, etc. This phenomenon is mostly known as a regular slower increase of animals' metabolic power compared to the increase of their mass. The effect was discovered in 19-th century by Rubner. Kleiber stated in 1932 two important properties of interspecific (across different taxa) allometric scaling, fascinating many researchers for the next 92 years. The first was that the metabolic rate $B$ mathematically is well described by a power function, in the form $B = aM^b$, where $a$ is a constant, $M$ is mass, and $b$ is the allometric exponent. The second result relates to the value of allometric exponent $b$, which Kleiber estimated of about 0.74, and - for convenience of presentation - rounded to 0.75 (3/4). This number became a benchmark, which other studies referred to since then. Such a power scaling was soon applied to other phenomena, which led to discoveries of many other similar allometric regularities observed with the increase of size of organisms or their constituents, including such as the density of pores in egg shells[1], biomass allocation to the seed coat in crops[2], gills in fish[3], etc. A detailed excurse on the subject can be found in many works, including Refs. 1, 4-7.

However, discovering the *fundamental* causes of MAS turned out to be a difficult problem. The consensus presently can be summarized as follows:

(a) Metabolic allometric scaling is due to cooperative action of multiple causes, both physiological, physical, environmental ones, and acting through natural selection[8,9], but not because of a particular physiological mechanism (which was a popular proposition for some time[10,11]).

(b) There is no a single universal value of the allometric exponent, nor a single deterministic basis common for all organisms, but different taxa may have substantially different allometric exponents, arising from different combinations of factors and mechanisms, defining this phenomenon[5,12-16].

As Ref. 9 notes, metabolic allometric theories usually consider physiological constraints arising from physical restrictions, needs for heat dissipation, nutrient influx and waste removal limitations, imposed by surface-area relationships, and natural selection as the main factors shaping organismal metabolism.

In earlier works[17,18], the author argued that the phenomenon of MAS is due to natural selection, but unlike in other studies, suggesting that such a selection and optimization of nutrient distribution occurs within the species forming a food chain and sharing common resources within the same habitat. In other words, it is the *entire food chain*, which is subjected to selection and optimization with regard to resource distribution among the species, forming the food chain. Of course, such selection occurs within boundaries, imposed by physical, environmental, physiological, organismal constraints, while these boundaries can change in time and space for different reasons. Solid proofs of validity of these author's ideas, including mathematical calculations and comparisons with experimental data, were presented and thoroughly analyzed. However, at that time, such ideas were outside mainstream paradigms. The situation changed for the better in recent years. The following quotations show how other researchers express similar idea of natural selection as the main cause of MAS.

In Ref. 8, the author asserts: "The selection of mass in multicellular organisms is therefore dependent on an interactive competition where mass is selected as a competitive trait that is used by the larger than average individuals to monopolise resources."

"These results suggest that mass-specific metabolism is selected as the pace of the resource handling that generates net energy for self-replication and the selection of mass".

"...it is by far primary variation in resource handling that generates the body mass variation and allometric scaling in mammals."

Overall, the author's of Ref. 8 position is that it is *natural selection*, which is a major player in defining mass and metabolic properties of organisms. The author also emphasizes separate consideration of energy required for replication, and how it can be linked to metabolism.

The author of Ref. 7 summarizes his findings as follows: "Metabolic scaling does not follow a universal physical law, but is a product of many contextual influences."

Another view, supporting the same idea, is formulated in Ref. 9 as follows: "... our approach expands the phenotypic space in evolutionary optimization operates and avoids giving primacy of causation to any single pillar of multicellular life. It also emphasizes that the pillars of *metabolism*, *growth*, and *reproduction* (italics by YS) have coevolved to shape each other, and, consequently, observed life-history strategies emerge from the optimization of these to maximize lifetime reproduction within a finite life span... Within this multivariate optimization dwells the great diversity of life histories in nature." Note that, as in Ref. 8, this work also emphasizes the role of reproduction as an important constituent of the overall energy expenditures.

We conclude this subsection adding that two types of MAS are distinguished: besides the *interspecific* allometric scaling, it is also considered ontogenetically or for the same species. In the last case, it is called *intraspecific* allometric scaling. The fundamental causes of the last one, according to Ref. 19, relate to cellular properties, modulated by heat dissipation abilities of organisms. Results from Ref. 19 show that mechanisms, defining interspecific and intraspecific allometric scaling, despite the similarity of names and observed effects, are rather different phenomena. We will deal with interspecific MAS.

## 1.2. *Phylogenetic correction and its flaws*

### 1.2.1. *Phylogenetic correction*

A note should be made about the so called "phylogenetic correction" (also called "phylogenetically informed approach"), whose idea is to first transform the *actual* data into a *virtual* space through complex mathematical procedures, using phylogenetic information[20-24], and then working with such data. The argument is that organisms' characteristics depend through the common phylogenetic history, and so such dependence has to be removed before one uses the data. Refr. 8-10, 25, 26 consider in detail, what kind of flaws this approach has, and why it cannot be universally applied, as its adherents insist. For instance:

(a) The approach accounts only for phylogeny, while there are lots of convincing evidences that ecological and other factors are at least of great influence too, and explain the observations incomparably better than the phylogenetically modified data (see the aforementioned studies).

(b) The phylogenetic approach itself and the obtained results generally are impossible to verify.

(c) Metabolism of living organisms, especially of multicellular ones, varies several times and often by orders of magnitude greater compared to accounted for phylogenetic features even for an individual organism, depending on many factors (current state - highly active, at rest, in torpor; season, phase of growth, age, food availability, etc.) Metabolism evolutionary had to be the most variable and quickly adjustable characteristic of any organism, capable to survive on the Earth.

(d) The main and rather the only claim of supremacy of phylogenetic approach is that modified data better fit linear regression curves. However, this is an *inherent* property of the used *mathematical procedures*, which will do the same for *any* data with positive correlation, regardless if they true or not[26]. The thing is that phylogenetically, traits are *always* positively correlated, so that, by and large, it does not matter, if the phylogenetic information is correct - the data fit will be *always* better. Despite this, the *values* of allometric exponents, obtained by phylogenetic approach, show higher divergence and irregularity, compared to conventional observations, which makes them indeed a very doubtful acquisition, unsuitable for making any constructive generalizations, even more so for solving the problem of interspecific allometric scaling.

(e) With regard to the whole body of all previous allometric studies, done almost for a century, this phylogenetic approach *invalidates them entirely*, in one gesture, which should not be the case, given so many important results obtained by predecessors with conventional approaches. Imagine Kleiber or Rubner used phylogenetic correction. Then very likely we won't be aware of a phenomenon of metabolic allometric scaling at all! One may object to this note that science develops iteratively and incrementally, and phylogenetic correction is next iteration. However, the thing is that this "iteration" refutes those original results, and so this is not *iteration*, but a *negation* of the mere foundation of the phenomenon.

So, this material does not use phylogenetic information, for which in Refs. 8-10, 25, 26, and in many other works, convincing proofs were presented. (Just a side note: application of phylogenetic approach even to a single discovered mechanism - from several, defining interspecific allometric scaling - produces absurdity; see Ref. 26.)

#### 1.2.2. *"Food chains" and "food webs" terminology in the context of the study*

A bigger animal can prey on several smaller animals, which in turn can be in dependent and independent relationships. Similarly, the same plant based food can be consumed by animals of different size. Smaller animals, like lions compared to buffalos, or wolves compared to moose, once organized in packs can feed on bigger animals as well. So, the picture of predator - prey relationships can be quite complicated; accounting for plant food adds even more complexity. When it was first introduced, the term "food chain" included this understanding, but then a more adequate in this regard term "food web" was supplemented, while in many instances the term "food chain" is still in use in its original wider meaning, especially in the popular literature. In this wider meaning the term is used in this paper, although with regard to the aforementioned contextual differences, the proposed study considers *food webs*, and the used methods and mathematical approaches are applicable to animals engaged in however complex relations within food webs.

In our case, the reason to resort to the term "chain" is that MAS is much tied to *mass* of organisms we study, and we consider metabolic properties of these organisms with relation to their *sequential* masses, as one of the main parameters. It is *this* sequence of masses, which in the context of our study should be addressed more appropriately as a "chain", rather than the "web" (of masses). Using chains of arranged by mass organisms, we thus introduce a great deal of order into our considerations, anchoring different aspects of the considered phenomenon to the same basis - sequential masses of organisms. Therefore, the term "food chain", used in this paper, should be understood in the context of sequential masses (belonging to not necessarily feeding on each other animals), but not as a rigid hierarchy of certain paired animals, tied to each other through predator - prey relationships.

The last impression could (wrongly) come from the appearance of mathematical approach we use, which considers energy expenditures for *two* animals with different masses. However, mathematically, there is no limitation, how much masses of animals can differ, neither that animals should relate somehow to each other through the food chain - the approach is as good for animals located at the opposite sides of mass spectrum, as for neighboring (by value of masses) animals. Allometric exponents here are not calculated based on relationships between *pairs* of animals, but for the *entire* ensemble of animals composing a given food chain. Of course, certain conditions to be fulfilled for such an approach can be justified; for instance, a linear dependence of metabolic rate on mass in logarithmic coordinates, small deviation of considered data points from the average value, absence of points-outliers. Note that these conditions in general case are not always fulfilled. Such, allometric exponents can be different at different stages of growth, usually being greater at the beginning - the effect, which is especially pronounced in growth of unicellular organisms[18].

## 2. Main concepts and methods of the study

This paper is much based on concepts and methods of the previous study[17]. We will start with exposition of the main ideas and results, to give a broader picture of the research.

> *Metabolic allometric scaling of considered multicellular organisms, which include mammals, birds, reptiles and fish, is a consequence of natural selection, which optimizes sharing of common resources of a habitat by organisms, which coexist there and form a food chain. It is a quantitative reflection of how different species within a food chain share common resources, in order to each species could continually reproduce, while at the same time being a sharable resource for the rest of the food chain. This arrangement preserves a delicate dynamic balance of distribution of habitat's resources, allowing perseverance in time and space of species composing the food chain, while avoiding overexploitation of common resources by one or several species to the detriment and even possible extinction of other species. The process of selection and optimization is framed by the boundaries defined by physical, environmental, physiological, organismal constraints at any given moment.*

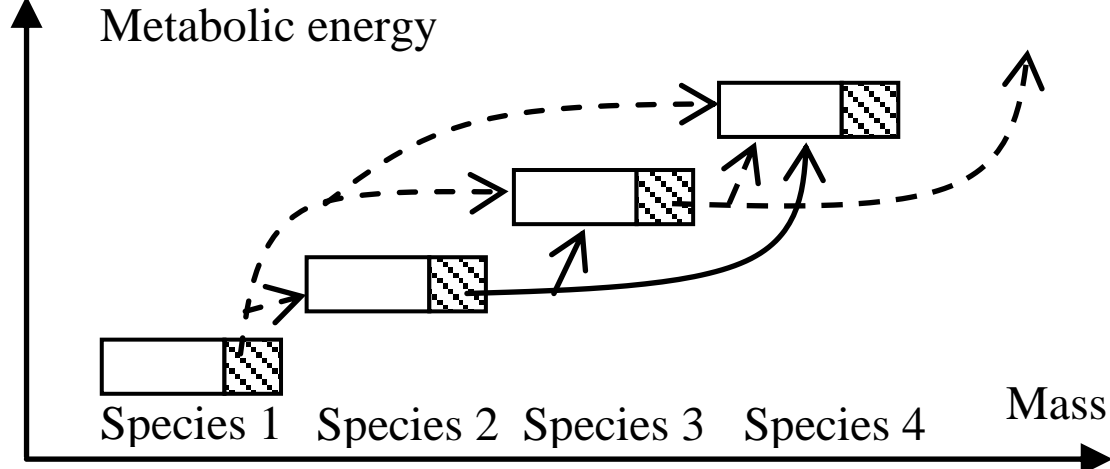

Fig. 1. Change of metabolic energy within the food chain of a habitat versus mass. Dashed rectangles present fraction of species' population consumed by the next (in mass hierarchy) levels of species. White rectangle denotes population that remains for reproduction.

Fig. 1 shows the concept in a graphical form. A dashed rectangular is presenting population consumed by species with a bigger mass at next levels (the same species can be consumed by species at several higher mass levels, like mice by cats and foxes). The remaining quantity of species population (a white rectangular) is sufficient to reproduce enough offspring to avoid extinction of this species. In such an equilibrium of a food chain, a metabolic rate is established as a product of selection in such a way that each next in the ladder species have sufficient resources, and as a consequence energy, supporting necessary functions and characteristics, like speed, to get enough food for population reproduction. On the other hand, this energetic and mass advantage is also restricted from the top, in order to not destroy the populations, the upper species prey on. If this happens, the current food chain balance is disturbed, and either the food chain could rearrange, or it is destroyed. This selection and optimization process evolves within the boundaries defined by physical, environmental, physiological, organismal constraints.

Arriving at such a conclusion is not straightforward. The subsections below outline the main considerations, which led to initial assumptions and obtained results, and which approaches and methods were used to prove the validity of original hypotheses.

#### 2.1.1. *Optimal resource sharing as a basis for existence of stable food chains within a common habitat*

Living organisms, existing within the same habitat, create food chains. This is an empirically confirmed phenomenon, which we can consider as an indisputable fact No. 1. Neither one can dispute the fact No. 2 that animals, comprising food chains, *share* common resources of their common habitat. The mere stable existence of a food chain in a given habitat is a proof of the fact No. 3 that this distribution of shared resources is *optimal*, in the sense that every species is getting sufficient resources for continuous reproduction, but not too much to jeopardize existence of other species within the food chain. Indeed, if the species are not getting enough resources, it will diminish in quantity, maybe even to the extent of extinction, thus affecting related to them through a food chain species, and producing misbalance of a food chain. If some species obtains too much common resources, its population will explode and will deprive other species of resources, possibly to the extent of destroying populations of other species, which will also disturb the balance of a food chain.

So, the essence of existence of any natural species (meaning the ones not overwriting natural instincts) is *resource acquisition* for reproduction. And, since species exist within food chains, this resource acquisition is performed through *sharing of common resources*, which in case of stable food chains is also *optimal*, in the sense of stability of a food chain discussed above. Thus, resource sharing is an *inherent property* (feature, a common denominator), with which existence of any species has to comply.

The acquired resources, in one way or another, are transformed by living organisms to *energy* (some resources can be stored for future use, of course, but they also present a source of energy). We can state one-to-one relationship between acquired resources and energy production, which is organismal metabolism. Therefore, sharing resources is equivalent to sharing energy produced from all consumed common resources. Sharing of resources, on the other hand, is optimal with regard to stable existence of a food chain, and so optimal, in the same regard, is sharing of produced metabolic energy by different organisms comprising the food chain. It was empirically found that this sharing of produced energy is performed in accordance with metabolic allometric scaling, for which allometric exponent is the main quantitative parameter in a mathematical description of dependence of produced energy on the mass of an organism. Therefore, metabolic allometric scaling is a quantitative representation of optimal distribution of common resources for energy production within a food chain.

Assembling all facts noticed before, and the conclusions above, into a logical causal sequence, starting from the fact No. 1 (of existence of food chains as such), one should come to the following *assumption*: Metabolic allometric scaling is shaped by natural selection, resulting in *optimization* of distribution of common resources between species composing food chains, while the metabolic allometric exponent appropriately represents a quantitative characteristic of an *optimal* distribution of energy between species, whose reproduction is supported by acquired shares of resources of their common habitat.

Then, the next question would be, how can one *prove* this assumption? However, before answering this question, we should understand, what sharing of resources mean, how it is actually performed in different food chains and environments, and how, by which virtues, it manifests itself.

### 2.1.2. *Sharing of common resources as a multifactor and multimodal phenomenon*

Sharing of common resources is achieved by animals using different means and combinations of means, playing to advantage of particular organisms. Among the most common are speed (to escape the predator and catch the prey), power (buffalos are slower runners compared to lions, but they are not an easy prey for lions due to their power and mass). Maneuverability is another important asset both in hunting and escape (recall videos, showing rabbits fleeing from a pack of polar wolves, birds escaping predators, birds catching insects, and so on). Endurance, ambush tactics, group hunting, certain evolutionary body and organ developments, and many other features form a vast toolset for acquiring a needed share of resources, exercised by different multicellular animals and unicellular organisms.

So, the overall panorama, by what combinations of means different animals acquire shares of common resources of a given habitat, is quite complex and diverse, and not rarely it is not obvious, how much different means contribute to the whole share of acquired resources. Such a multifactor situation makes direct calculation of allometric exponents on the basis of acquired resources difficult, but not impossible, especially when one or few dominant features are used, whose characteristics can be accurately quantified. Animal speed is one of such characteristics. When less options for resource acquisition is available, the importance of speed appropriately increases, and manifestation of speed in this capacity should become more prominent. This consideration is supported by the results of different studies. For instance, in Ref. 27 the authors acknowledge: "Terrestrial mammals inhabiting structurally simple habitats tended to be faster than those in complex habitats." Indeed, simpler habitats offer fewer possibilities for development of other means of acquiring resources, complementing speed advantage; hence, a greater reliance on speed, and accordingly a greater share of resources (and energy) devoted to support this feature, and, with regard to our needs, the more adequate and accurate will be calculations of allometric exponents based on speed advantage (since in this case the input of other unaccounted factors will be less).

Speed advantage is a widely used, and often the major and the most critical one, means for resource acquisition, and in many instances it can be measured with sufficient accuracy. So, here, we consider speed as a manifestation of energetic abilities of species, supported by the species' share of acquired resources. In many instances, speed is an adequate *integral* characteristic of the *overall* energetic advantage. Of course, an energetic advantage can manifest itself in different ways, and as a combination of different characteristics. However, the more dominant (with regard to resource acquisition) and resource demanding some feature is, the more adequate will be calculations of metabolic parameters based on this feature. The author thinks that this idea of energetic advantage, expressed in different ways, is a rather general phenomenon and maybe a universal criterion applicable to many taxa. Maybe even to plants.

Saying this, a note still should be made that animals generally use *combination* of different means to acquire resources, and the issue of how much each "supply channel" contributes to resource acquisition is of importance, since ideally each contributing fraction of resources should be taken into account, or at least to be evaluated as insignificant factor. However, this is not always possible. Our analysis based on speed is not exception, and besides the speed other factors, affecting resource acquisition, present - in different degree for different classes of animals. We made selection of animals for which speed is the defining factor in resource acquisition, but still some influence of other factors (different in different classes of animals) remains in this selection. In this regard, in Appendix D.3 we consider birds as an illustrative example from our study, how other factors, besides speed, may affect metabolic allometric scaling and accuracy of our approach.

#### 2.1.3. *Methods*

*Proof of validity*

Now we are ready to answer the question asked at the end of subsection 2.1.1: How can one prove the assumption that the metabolic allometric scaling is shaped by natural selection resulting in optimization of distribution of habitat's resources between species composing its food chain? For that, one should find the amount of resources (or values of some closely, one-to-one, related to resources characteristic) required to support metabolism of an organism with a certain mass within a particular food chain, and then calculate the value of allometric exponent for these organisms based on the values of found resources (or related to them characteristic). Then, we should compare results of such calculations with experimentally found values of allometric exponents for the same (ideally) or similar species. If calculated values match experimentally found allometric exponents, then our assumption is likely true. Coincident match is possible, of course. However, if we apply the same approach to different classes of animals, and in all instances obtain a good match of calculated and experimental data, then the probability that our assumption is true greatly increases.

This approach follows from the general principles of validation of scientific hypotheses. The following considerations can be also used.

(a) The closer calculated numbers and experimental data match, the greater is probability that a hypothesis is true. (Our calculated data in most instances match experimental results very accurately, which makes probability of accidental coincidence extremely low; in practical terms - actually zero.)

(b) The more representative and more reliable are ensembles of used data, both for calculation of theoretical values, and of experimental data for comparison, the greater is probability that the hypothesis is true.
(c) Results of previous studies, reported in the literature, support our hypothesis. For instance, there were studies showing that incorporation of allometric scaling into the dynamic equations of food web models enhanced their *stability* (italics by YS), when allometric scaling was used to relate masses of preys and predators. Our assumption considers stability of food chains as a consequence of optimized distribution of resources, described by allometric scaling, and the aforementioned fact points in the same direction.

The situation with our results with regard to their validation is typical for any non-trivial discovery, whose validity hypothetically can never be hundred percent true, but in practice is considered as a proven knowledge, once the combination of criteria and characteristics, used for validation, summarily exceeds a certain threshold. (Some pedants still assign probability $10^{-7}$ that Newton's Second Law of mechanics is invalid, while a great deal of technical might of civilization is already built upon this physical law.)

*Advantages using integral parameters*
In most cited above references, such as Refs. 5-9, the authors prefer considering different constituents of MAS phenomenon, such as metabolism, growth, reproduction, etc., and then assembling knowledge about the phenomenon on the basis of properties of studied components. However, the results will be more reliable and natural if we can consider the *whole* metabolic energy, and then move *downward* to more particular knowledge and composition of constituents. This is the approach used in this paper, associating the total energy with the integral parameter speed, so that we do not need to make assumptions about distribution of the total metabolic energy between different organismal constituents and functions. (In case of birds, however, we consider possibility adding characteristics requiring additional metabolic energy to explain lower value of maximal metabolic rate.) Certainly, the more speed is critical for resource acquisition, the more adequate and accurate calculations of allometric exponents based on speed will be.

Since the basis for our comparison of metabolism of different species is animals' speed, we developed biomechanical models. Using them, we find allometric exponents based on differences in speed, mass, skeletal and body compositions, like limbs. Then, we compare obtained numerical results with known experimental values. For the species with representative datasets the correspondence between calculated and experimental data was very good. Given the fact that values of allometric exponents used for comparison were obtained *independently*, by different researchers, and on very different grounds, this is a strong indication of validity of obtained results and the main hypothesis.

### 2.1.4. *A notion of evolution*

For the following, we need to specify the notion of evolution in the context used in this paper, because different researchers might assume somewhat different contexts this term means for them.

Interspecific allometric scaling in animals, we argue, is the result of simultaneous working of several factors, evolving within the context of natural selection. Action of biomechanical constraints is one of them. Its understanding requires knowledge of mechanics. The other important factor relates to physiological and evolutionary principles, which is a more subtle issue. The notion of group selection is an example of such a controversy. For that reason, we have to make a small excurse to the modern evolution theory[27]. This work states the foundational concept of evolution as follows: "The laws of physical science plus natural selection can furnish a complete explanation for any biological phenomenon". The overall attitude to group selection in Ref. 27 is summarized in the statement: "Benefits to groups can arise as statistical summations of the effects of individual adaptations". These two ideas, in fact, form the conceptual foundation of this study too.

This stance has to be made clear, because some researchers are very likely to raise a "red flag", once they read about directed evolution of species within the food chain. As Williams, the author of Ref. 27, we are also talking about "statistical summations of the effects of individual adaptations", caused by the same environmental conditions, affecting each species *individually*. (Of course, with a possibility of a certain transfer of acquired features to offspring.)

In this work, we show that fundamental causes of interspecific allometric scaling originate as the result of adaptations of species *within the food chain*, under the influence of similar factors. One can call such a phenomenon as a *systemic evolution* or as a selection's adaptation of an entire food chain, carefully stipulating what evolution and its numerous aspects mean in this context. However, our purpose is to show real physiological and evolutionary mechanisms, defining so far mysterious phenomenon of interspecific allometric scaling, while the inviting generalizations of obtained results and philosophical issues are left for the future, if any.

## 3. Biomechanical models of animals for MAS studies

### 3.1. *Selection and adaptation of animals within the food chain*

This study of interspecific allometric scaling began in search for particular physiological mechanisms, which collectively could define this phenomenon. However, in the process of studying this problem, the author gradually came to a conclusion that this is probably the interaction of species within the food chain and with the environment they act in to acquire resources, which shapes their metabolic properties. This interaction, subjected to natural selection and optimization, regulates the "appetites" of creatures composing a food chain, and accordingly their metabolic properties and mass.

In normal conditions, the food chain itself has properties of continuity and dynamic balance. The metabolic energy of an organism, belonging to a food chain, must be sufficient to acquire enough nutrients for a successful reproduction and maintenance, but not excessively strong to jeopardize the reproduction of species the organism directly feeds on or indirectly interacts with, like sharing common resources, within the food chain.

Since environmental conditions change all the time, such a balance, by its nature, is a dynamic one. In an established state, the food chain is continuous. Indeed, whatever species we take, we can associate with it a list of "who eats who and what" in both directions (of course, one animal can feed on several animals, and use plant based food sources, so the relationships can be quite complex). When a food chain is broken, the organisms within it tend to "repair" the damage and restore the continuity through certain transitional steps, establishing new balance. The proofs for these statements, now not obvious, will be provided in a due course. In support, let us quote Ref. 14, which also mentions "the necessity to share resources with competitors", thus also underlying the importance of considering common for the habitat resources, shared among different species.

The second important property of living organisms is the following. Evolutionary development of organisms equipped them with adaptation capabilities, such, that using combination of different physiological mechanisms and developing new ones, they can adapt to a *very* wide range of environmental conditions, far exceeding limitations imposed by particular physiological mechanisms. Such mechanisms are the *means* serving the main purpose of any species - survival for the reproduction. Mechanisms can be enhanced, combined in different ways, new ones can be developed, but these are not the mechanisms, which define the limits of evolutionary development, but the need for the reproduction of species in conditions, imposed by environments, so that organisms mobilize all possible resources for this purpose.

Organisms living in the same habitat, within the reach of each other, eventually create a single food chain. Even if species do not feed directly on each other, they share common nutritional environment, like vegetation, seeds, fruits, nuts, even common atmosphere contributes to their linking. The thing is that the links in the food chain are not straightforward. Such, cats feed not only on mice, but, nonetheless, evolutionarily cats developed in such a way that they cannot overexploit ability to catch their preys, so that the preys could reproduce in sustainable numbers, for the benefit of their own but also for the cats.

Examples of high adaptability of living organisms are numerous. Even humankind presents extremely high variability of all characteristics, including metabolic rates. Athletic training and world records is one example, while the great diversity of human populations adapted to different geographical zones and sometimes very specific habitats is another example. Whatever exotic characteristics the environment has, there are almost surely some living organisms finding their "home" there. Each geological period on the Earth, which provided some minimal conditions, had some life forms. So, we may consider this as sufficient evidence that no single intrinsic factor could fundamentally limit organismal adaptation, save for some extreme conditions. In Ref. 28 a similar idea is formulated as follows: "evolution is constrained by physical laws, but … the effects of these laws can be modified by biological innovation". (Illustrative animated locomotion optimization can be seen at goatstream.com/research/papers/SA2013/index.html.) The fact that multicellular organisms exhibit a range of allometric exponents, depending on the physiological regime, also supports the thesis that living organisms can adjust their metabolism to very different environmental conditions through diverse organismal constitutions[25,12-14,15].

### 3.2. *Biomechanical model. Finding its attributes*

There are many studies devoted to development of biomechanical models of animals' locomotion. Recent Ref. 29 presents a rather typical rigid pendulum model of leg movement, modeled by pendulum in swing and an inverted pendulum in stance. The work also provides a list of the literature on the subject. In Refs. 1, 30, the scaling of bio-mechanical constraints was considered, such as mechanical capacity of limbs to withstand buckling and pressure. Other works studied geometrical, kinematic and dynamic mechanical parameters of organisms and their scaling relationships[31,32]. However, such studies, undoubtedly very useful, did not shed light on the fundamental level mechanisms defining metabolic allometric scaling. In some instances, the discovered scaling patterns do not match the results predicted by models. Such, in Ref. 32, the authors acknowledge that "limb inertial properties do not have the potential to underlie cost of transport (COT) scaling". Such a relationship, if it existed, could optimize the energy expenditure, but it is not the case. This means that other organismal demands, which are more important, override such optimality. Similarly, other scenarios, considered in the same work, confirm that other than merely mechanical optimization factors have more impact on organisms' development.

We will consider mechanical factors, which most closely relate to metabolism, that is energetic and kinematic ones. In particular, when the speed is of primary importance for the survival of species (which is true for many organisms), this means that a predator and a prey have to have commensurate speeds, with some advantage on the predator's side.

#### 3.2.1. *Horizontal motion of limbs*

First, we consider the *minimal* mechanical energy requirements for motion, assuming that the predator and the prey move with the *same* speed. This will give us the base allometric exponent, to which the components due to other factors, such as a certain speed advantage, will be added later. Such decomposition turned out to be an efficient approach.

Speed is achieved through the motion of limbs. Fig. 2a presents their rotational motion. However, when displacement $S$ is small compared to the limbs' length, mechanically, the translational motion of a center of mass is an accurate approximation.

Bodies of both animals move with the same speed $V_b$ (the upper ends of limbs). The fractions of time spent on moving limbs forward and backward are the same for both animals. Then, the lower ends of limbs of both animals, when moving forward, have to have the same average forward speeds $V_{avf}$ in order to support equal average velocities of bodies. A bigger limb moves from the position 1 to the position 2; a smaller limb from the position 3 to 4; the apex angles are the same. Limbs have accordingly lengths $L$ and $l$, masses $M_l$ and $m_l$. The centers of masses of both limbs have average forward velocities $V_{cf} = (V_b + V_{avf} c_m)$. Here, $c_m$ is a fraction of the limb length, measured from the point of limb's attachment to the body, corresponding to location of a center of mass (COM). (In Ref. 32, it was estimated to be of about 1/3.) This location of COM for mammalians, says this work, scales according to geometric similarity, that is remains unchanged. The authors acknowledge: "For all subgroups, fore- and hindlimb COM position scales according to geometric similarity, … indicating that limb mass distribution remains unchanged with respect to increasing body mass."

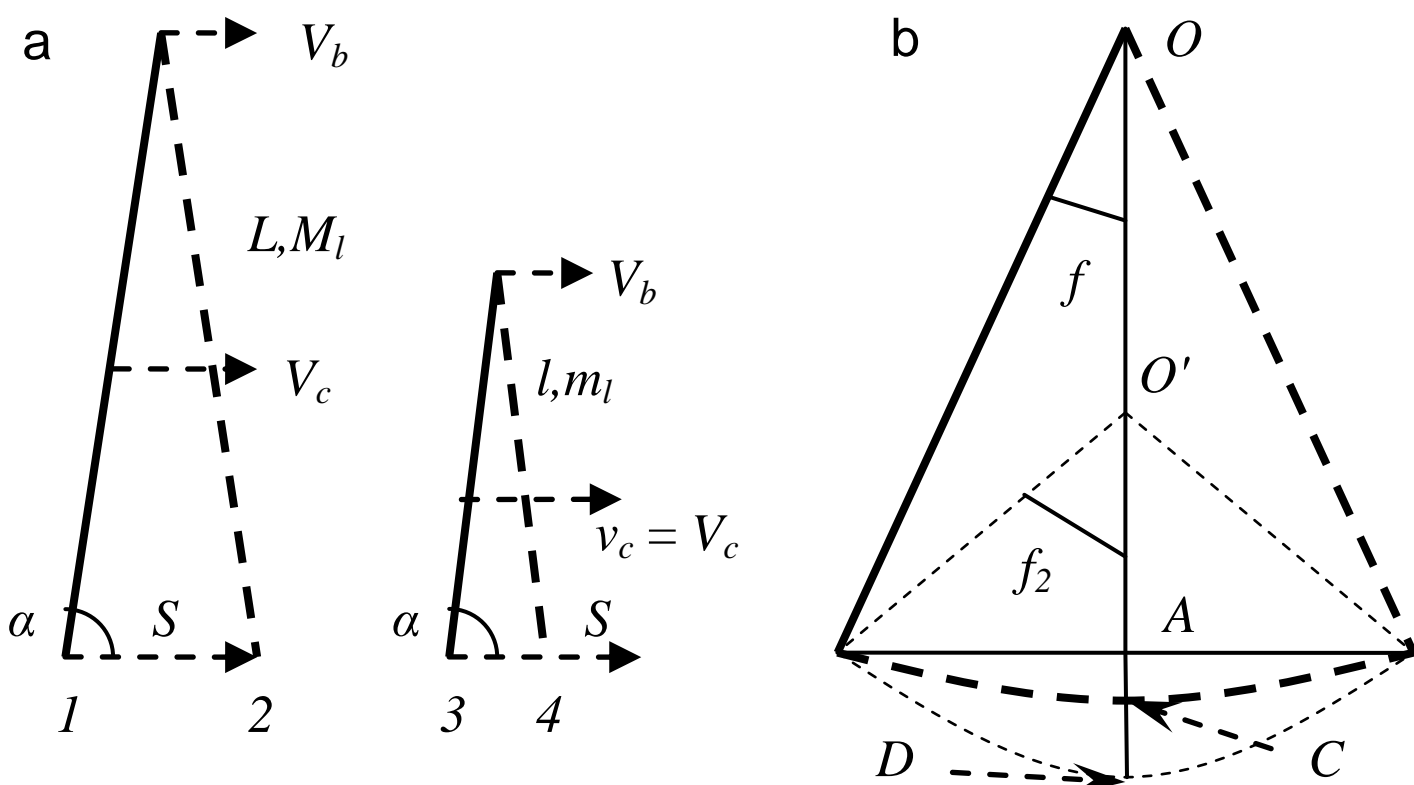

Fig. 2. Forward and backward motion of animal limbs. a - Movements of bigger and smaller limbs. b - Calculating differences in potential energies required for the vertical motion of bigger and smaller limbs. Strides are equal.

Similarly, the average backward velocities of centers of limbs $V_{cb}$ are equal too.

We want to know the ratio of energies, accordingly $E$ and $e$, required to move limbs by the distance $S$ for the same time (which, obviously, means the same speed). For the forward movement, we should take into account the increase of velocity of centers of masses from the backward to forward velocities, which, as we found, are the same for both animals and equal to $V_c = (V_{cf} - V_{cb})$. Then, the ratio of kinetic energies (in the bodies' systems of coordinates) is as follows.

$$\frac{E}{e} = \frac{M_l V_c^2 / 2}{m_l v_c^2 (L/l)/2} = \frac{M_l}{m_l} \times \frac{V_c^2}{V_c^2 (L/l)} = \frac{M_l}{m_l} \times \left( \frac{l}{L} \right) \quad (1)$$

If we assume geometric similarity, that is $(l/L) \propto \left( m_l / M_l \right)^{1/3}$, then Eq. (1) transforms to the following.

$$\frac{E}{e} = \frac{M_l}{m_l} \times \left( \frac{l}{L} \right) = \left( \frac{M_l}{m_l} \right)^{2/3} \quad (2)$$

or

$$E = e\left(M_l / m_l\right)^{2/3} \qquad (3)$$

In other words, in order to support the same speed, the bigger animal needs more energy proportionally to the ratio of limbs' masses in the 2/3 power, which is the value of the "mechanical" part of an allometric exponent in case of geometric similarity.

In fact, the ratio $(l / L)$ scales differently for different groups of animals, which, as we will find out, is one of the reasons why different groups of living organisms have different allometric exponents in the same physiological state.

The term ($l/L$) in Eq. (1) takes into account that in order to cover the same distance $S$ for the same time, the smaller animal needs to make each step faster and has to make $(L / l)$ times more steps.

The notion of geometric similarity includes both proportional increase of limbs' lengths, and also increase of limbs' mass proportionally to mass of whole organisms. The last assumption is fulfilled with high accuracy for mammals[32]. The authors conclude, "Across quadrupedal mammals, limb mass scales isometrically with body mass". (Note that our approach allows taking into account insignificant positive allometry of 1.01 and 1.03 for fore- and hindlimb mass increase discovered in Ref. 32, if necessary.) The assumption about 3-D increase of limbs is also supported with reasonable accuracy by studies of body proportions to weight in different animals, reviewed in Ref. 30.

Similarly, we can find the ratio of energies during the backward movement of limbs. In this case, the limbs push off the supporting surface with the force proportional to mass. Indeed, both bodies have the same speed $V_b$, which is supported by providing about the same acceleration $a$ at each push off. (This proposition, of course, can be explored in detail, in order to find out the second order adjustments for particular geometry and mode of motion. However, here, we restrict ourselves to the first order approximation, for whose validity there are many mechanical reasons, due to similar dynamics and geometry of motion.) For such assumptions, the force $F = a\mu$, where $\mu$ denotes a generic mass. As we did above, we want to find the energy consumption when both animals cover the same distance $S$. Then, the ratio of energies is

$$\frac{E}{e} = \frac{aMS}{amS(L / l)} = \frac{M}{m} \times \left(\frac{l}{L}\right) \qquad (4)$$

If we assume geometric similarity, then Eq. (4) transforms into

$$\frac{E}{e} = \frac{M}{m} \times \left(\frac{l}{L}\right) = \left(\frac{M}{m}\right)^{2/3} \qquad (5)$$

(Thus, we again obtained the allometric exponent of 2/3.) In Eq. (4), the term ($l/L$) accounts for the fact that the smaller animal has to make more steps - same as in Eq. (1). Note that we deal with the whole body masses $M$ and $m$. The bones of a bigger animal might be disproportionally heavier, but we will account for this factor later.

Now, let us consider the forward movement of limbs as rotation. In recent years, this approach received attention, since it was discovered that energetic costs of limbs' swinging comprise 8 to 33% of the total locomotor costs[32,33]. In our case, we are interested in the ratio of kinetic energies, which is as follows:

$$\frac{E}{e} = \frac{I_M \Omega^2}{I_m \omega^2}\left(\frac{l}{L}\right) = \frac{M_l L^2 \Omega^2}{m_l l^2 (\Omega L / l)^2}\left(\frac{l}{L}\right) = \frac{M_l}{m_l} \times \left(\frac{l}{L}\right) \qquad (6)$$

Here, $I$ denotes moments of inertia of rotating limbs, $\Omega$ and $\omega$ are angular speeds, accordingly of the bigger and smaller animals. Eq. (6) accounts for the fact, that, since the apex angles are the same for both animals, the smaller animal has to exercise a greater angular speed by $L/l$ times, in order to provide the same horizontal speed of body.

In case of geometric similarity, Eq. (6) produces allometric exponent of 2/3.

$$\frac{E}{e} = \frac{M_l}{m_l} \times \left(\frac{l}{L}\right) = \left(\frac{M_l}{m_l}\right)^{2/3} \qquad (7)$$

Similar to (6) and (7), we can consider the backward rotational motion of limbs (accounting for the fact that a smaller animal has to make $L/l$ times more steps). In this case, the moment of inertia will include both the limbs', and a part of body's mass (since in quadrupedal animals two limbs can work simultaneously). This consideration will not change the final ratio, since, given the aforementioned result from Ref. 32 about isometric scaling of limbs' mass, we can substitute instead the mass of whole animals (when two limbs are involved in a backward movement, like in a gallop) or equally apportioned parts of whole masses.

So, the ratios of kinetic energies in both modes of motion are described by the same mathematical expression $(M_l / m_l) \times (l / L)$ (or $(M / m) \times (l / L)$, if we account for isometric scaling of limb mass). In case of geometric similarity, these expressions produce the allometric exponent of 2/3.

### 3.2.2. *Vertical displacements*

Let us consider the situation when the apex angles are different, while both animals make equal steps (Fig. 2b). The striding angles are accordingly $2f$ and $2f_2$; vertical displacements required for limbs to not touch the surface are $H = AC$ and $h = AD$; arcs represent parts of circles with radii $L$ and $l$ and with centers at points $O$ and $O'$. Then

$$H = L(1 - \cos(f)) \approx Lf^2 / 2 \qquad (8)$$

$$h = l(1 - \cos(f_2)) \approx l(1 - \cos(fL / l)) \approx l(fL / l)^2 / 2 \qquad (9)$$

Here, we used the first two terms of the Taylor's series representation of cosine. Using (8) and (9), the ratio of potential energies, required to overcome the force of gravity, is

$$\frac{M_l g H}{m_l g h} = \frac{M_l}{m_l}\left(\frac{l}{L}\right) \qquad (10)$$

In case of a geometric similarity, the allometric exponent $b$ is equal to 2/3.

$$\frac{M_l gH}{m_l gh} = \frac{M_l}{m_l}\left(\frac{l}{L}\right) = \left(\frac{M_l}{m_l}\right)^{2/3} \qquad (11)$$

So, the ratio of potential energies is the same as for kinetic energies.

However, when the apex angles are equal and steps have different lengths, the allometric exponent is equal to one, $b = 1$:

$$\frac{M_l gH}{m_l gh} = \frac{M_l L(1-\cos f)}{m_l l(1-\cos f)(L/l)} = \left(\frac{M_l}{m_l}\right)^{1} \qquad (12)$$

Here, we took into account that a smaller animal makes ($L/l$) more steps.

The comparison of potential energy versus the kinetic energy for particular motion scenarios shows that the energy of vertical displacements is relatively small compared to kinetic energy. Besides, real animals significantly reduce vertical displacements, compared to our model, by bending limbs in joints. Videos showing chasing and escaping quadrupedal animals demonstrate that vertical oscillations of animals' bodies are very small. From the evolutionary perspective, since the vertical oscillations require additional energy, it makes sense to minimize them when possible, which, apparently, was the evolutionary path the development of animals followed. So, the energy expenditures on vertical oscillations are at least several times less than the ones for horizontal movements, while theirs base allometric exponent, given about the same striding angle, is the same as for horizontal motion (see Eq. (10) vs Eqs. (1), (4), and 6)).

Note that other more complex forms of motion can be always decomposed into combination of rotational and translation movements, which we considered. So, the obtained formulas have more general appeal than only for description of particular motion scenarios, and can be used as a basis for more sophisticated mechanical modeling. However, as we will see, the introduced models provide an adequate mathematical description of the phenomenon for our purposes and explain the causes of interspecific allometric scaling.

### 3.2.3. *Proportionality to velocity of energy expenditures for moving*

This is an important subject for our studies. As some comments showed, it is not understood well. The kinetic energy is proportional to *square* of velocity, which misled some people to think that the energy expenditures for moving should be also proportional to *square* of velocity. In fact, this is not so. It was found in Ref. 34 experimentally that the velocity of animals is *proportional* to used energy, although the authors could not explain, why it was so. This fact was also discussed, with some surprise, in Ref. 1.

Here is a theoretical explanation of this interesting observation. Let us compare energies required for an animal to move its limbs with different speeds $V_1$ and $V_2$ due to different lengths of strides $s_1$ and $s_2$ (for certainty, we assume $s_2 > s_1$), while making strides for the same time $T$. The animal has mass $M$.

As we discussed already, motion consists of acceleration and deceleration at each step (to a much lesser extent of the body, and more noticeably of limbs, whose velocity at the moment, and at the point, of contact with the ground is zero). We consider bodies, whose speed, obviously, is affected by acceleration and deceleration of limbs too, according to Newton's Third law of mechanics. Let us denote such change of speeds $\delta V_1$ and $\delta V_2$. Since we consider the same animal, we can assume the geometric similarity of motion (except for the length of strides), which accordingly entails linear scaling of dynamic forces and kinematic characteristics of motion. In particular, this means that the speed increments $\delta V_1$ and $\delta V_2$ relate to each other as speeds themselves, that is $\delta V_2 / \delta V_1 = V_2 / V_1$. On the other hand, since strides are made for the same time *T*, $V_2 / V_1 = s_2 / s_1$, and, consequently, $\delta V_2 = \delta V_1 (s_2 / s_1)$.

Animal's body moves in the range of speed $(V_1, V_1 + \delta V_1)$, with $\delta V_1 << V_1$. The change in energies, corresponding to speeds $(V_1 + \delta V_1)$ and $V_1$, during one step when the animal covers distance $s_1$, is as follows.

$$E_{1,s1} = M(V_1 + \delta V_1)^2 / 2 - MV_1^2 / 2 = (M/2)(2V_1 \delta V_1 + \delta V_1^2) \approx MV_1 \delta V_1$$

Here, we took into account that $\delta V_1 << V_1$, and consequently $2V_1 \delta V_1 >> \delta V_1^2$.

Accordingly, to cover distance $s_2$, the animal has to spend energy

$$E_{1,s2} = MV_1 \delta V_1 (s_2 / s_1)$$

Similarly, to cover distance $s_2$ when moving in the range of speed $(V_2, V_2 + \delta V_2)$, the animal has to spend amount of energy

$$E_{2,s2} = MV_2 \delta V_2 = MV_1 \delta V_1 (s_2 / s_1)^2$$

Then the ratio of spent energies is

$$\frac{E_{2,s2}}{E_{1,s2}} = \frac{MV_1 \delta V_1 (s_2 / s_1)^2}{MV_1 \delta V_1 (s_2 / s_1)} = \frac{s_2}{s_1} = \frac{\delta V_2}{\delta V_1} = \frac{V_2}{V_1} \tag{13}$$

One can also consider change of energy of limbs using rotational model, with the same assumptions as above, and calculating energies of rotating limbs through moment of inertia, similar to what we did in Eq. (6).

Effectively, Eq. (13) states that the amount of required energy, when the speed increase is due to longer steps, indeed, is proportional to animal's speed (but not to the square of speed), which was found in Ref. 34 experimentally. One can argue that different body parts have different accelerations. However, as Eq. (13) shows, what matters is the *ratio* of velocities, which is the same for limbs and body due to the geometric similarity of motion. Thus, (13) describes motion of the whole animal, moving at different speeds.

With regard to animals of different size, our considerations are applicable to different animals too. The authors of Ref. 32 mention that the stride angles in bigger and smaller animals are close, which means that bigger animals, indeed, make proportionally bigger steps. In case of more complicated motion scenarios (like different time for steps), the result will be the same.

Let us prove that the used energy is proportional to velocities for *potential* energy too, using a vertical ascent scenario. (It can be generalized for other forms of motion and for more factors.)

The potential energy *E* required to ascend the height *H* is equal to $E = mgH$, where *m* is mass in *kg*, *g*=9.81 $m \cdot s^{-2}$ is the acceleration of a free fall. If a climber is able to develop power *W*, then the ascending time *T* will be $T = mgH / W$. Velocity *V* is equal to $V = H / T = W /(mg)$, that is $V \propto W$. So, indeed, velocity is proportional to power in this case too (to the amount of energy spent per unit time).

Thus, we theoretically proved the result, obtained experimentally in Ref. 34, that the energy costs for motion are *proportional* to velocity. We did not take into account the air resistance and other possible secondary factors; however, in case of more detailed studies, they can be accounted for too.

3.2.4. *Scaling of velocity and other characteristics with mass increment*

We will need a mathematical method for finding allometric exponents for speed increase with relation to mass increase. This suggestion implicitly assumes that the speed, similarly to metabolic rate, also changes as a power function. Ref. 32 actually confirms this: "physiologically equivalent speeds are positively allometric with body mass, scaling approximately as $M^{0.21}$". So, we can use a power function for the velocity increase too.

Let us consider change in velocity when each next evolutionary stage produces a bigger animal with a greater speed. The rationale is that the bigger organisms could originate because of the predispositionally higher metabolism, compared to competitors. Once organisms become bigger, they have to maintain higher metabolic capacities, in order to reliably acquire nutrients for reproduction. For instance, in mammals, such an advantage in many instances is transformed into increase in maximal speed for each successive developmental phase (which is confirmed by the aforementioned result about the positive allometric exponent of 0.21 for velocity[32], and we will present more data on this account too).

Let us consider *x* number of hypothetical evolutionary development stages, each producing a bigger organism with mass $M_x$. The relative mass increase is by *g* times at each stage.

$$M_x = M_0 g^x \tag{14}$$

where $M_0$ is the initial mass. The velocity increase is $g^{b_v}$ times per development phase ($b_v$ is the allometric exponent for velocity, like the value of 0.21 mentioned above). Then, the velocity $v_x$ at *x* phase is as follows.

$$v_x = v_0 g^{b_v x} \tag{15}$$

where $v_0$ is the velocity at the initial stage.

Substituting the value of $g^x$ from Eq. (14) into (15), we obtain

$$v_x = v_0 (M_x / M_0)^{b_v} \qquad (16)$$

In other words, the allometric exponent for velocity $b_v$ does not depend on the number of developmental stages, nor on the mass increment $g$.

The solution of (16) is as follows.

$$b_v = \ln(v_x / v_0) / \ln(M_x / M_0) \qquad (17)$$

Below, we assume that the density is constant, so that mass is proportional to volume $U$, $M \propto U$, and consequently $M_x / M_0 = U_x / U_0$.

Note that the obtained Eqs. (16) and (17) are valid for *any* other parameter, whose change is associated with mass (or other reference value) and can be described by a power function.

### 3.2.5. *Locomotion as a factor shaping metabolism of whole organisms*

During the motion of limbs, other than mechanical energy expenditures occur (average muscle efficiency is in the range of 20 - 35%). How does this fit into our approach? This is where we come to an important consideration. In organisms, which rely on motion to acquire nutrients, locomotion is the *primary* function supporting organisms' existence, regardless of the motion mode, like by virtue of limbs, flagella, fins, tail in fish, body movements, etc. Even if animals use ambush tactics, at the end they have to overcome a prey in direct contact. Of course, there are many rather sedentary organisms, whose metabolic capacities are shaped differently. For instance, our separate study considered metabolic properties of unicellular organisms, for which the size and associated ability to acquire nutrients through the surface are the major factors[18].

For animals relying on movement, locomotion is the *main* function, the *reference base*, to which metabolism of organisms is adjusted evolutionarily and ontogenetically, so that the MAS defined by locomotion function should be propagated *through the entire organism*. This is a very natural arrangement, which stems from the most important and literally vital need to secure population reproduction by mobilizing all resources, while preserving a dynamic balance and continuity of the food chain (otherwise, there will be no source of nutrients). So, although the found allometric exponent is based on mechanical energy required for motion, biochemical metabolic activities supporting this energy production have to adjust to the *same* scaling. (When other factors play a major role, then evolutionary paths become different.)

### 3.2.6. *Accounting for the increase of skeleton mass*

It was shown in Ref. 35 and discussed in Ref. 1 that the mass of mammals' skeletons scales as the total mass at a power of $1.08 \pm 0.04$, so that the allometric exponent for the skeleton weight is $b_{sk} = 1.08 \pm 0.04$. This fact has important implications for the metabolic rate. Recall that endurance athletes increase their physical capacities not so much through the weight gain, but mostly through the boosting the metabolic capacities of existing muscles and supporting physiological mechanisms and systems. The same situation happens with animals when their skeletal mass increases. In order to better understand this, Appendix A presents an example with supporting considerations. It follows from this example that the increase of metabolic power has to be proportional to the weight load. It means that the metabolic gain should be proportional to the skeleton weight increase, and consequently scales in the same way. For instance, this consideration will change Eq. (6) as follows.

$$\left(\frac{E}{e}\right)_{sk} = \left(\frac{M_l}{m_l} \times \left(\frac{l}{L}\right)\right)^{b_{sk}} = \left(\frac{M}{m}\left(\frac{l}{L}\right)\right)^{b_{sk}} \tag{18}$$

Here, we accounted for the earlier discussed fact that the mass of limbs scales isometrically with the body mass[32].

Previously, we discussed that metabolism of moving organisms we consider is adjusted to the most important organismal function for the survival - to motion, which means that the energy for mechanical motion is proportional to the total energy produced by the organism. Thus, Eq. (18) is valid for the metabolic output of whole organisms too.

#### 3.2.7. *Calculating scaling of metabolic rate required to support increase of animal speed*

The allometric exponent which we considered so far represented the *minimum* value, which corresponds to *equal* velocities of a predator and a prey, while in order to catch the prey, the predator needs greater velocity, and accordingly higher energy. How much greater the velocity of a predator should be? One of the important variables is the *relative* speed of a predator compared to the prey. We will estimate how the value of allometric exponent $b$ above some threshold value $b_t$ translates into the speed advantage, assuming that the energy increase $\Delta E = E_2 - E_{2t}$ is proportional to velocity, as we found previously. Index '$t$' denotes threshold values.

Below, indexes '1' and '2' correspond to smaller and bigger animals.

$$\Delta E = E_2 - E_{2t} = E_1\left(\frac{M}{m}\right)^{b} - E_1\left(\frac{M}{m}\right)^{b_t} = E_1\left(\frac{M}{m}\right)^{b_t}\left(\left(\frac{M}{m}\right)^{b-b_t} - 1\right) \tag{19}$$

We can rewrite (16) also as $E_{2t} = E_1 (M/m)^{b_t}$ , since velocities are proportional to energies. Then, (19) transforms to the following.

$$E_2 / E_{2t} = (M/m)^{b-b_t} \tag{20}$$

On the other hand, the left part of (16) can be presented as the ratio of velocities (as we proved previously). Since the threshold velocities of large and small animals are equal, that is $V_{2t} = V_1$, we find.

$$V_2 / V_1 = (M/m)^{b-b_t} \tag{21}$$

Fig. 3 shows the graphs of relative velocity increments versus the relative mass increase depending on increments of allometric exponent, calculated using Eq. (21).

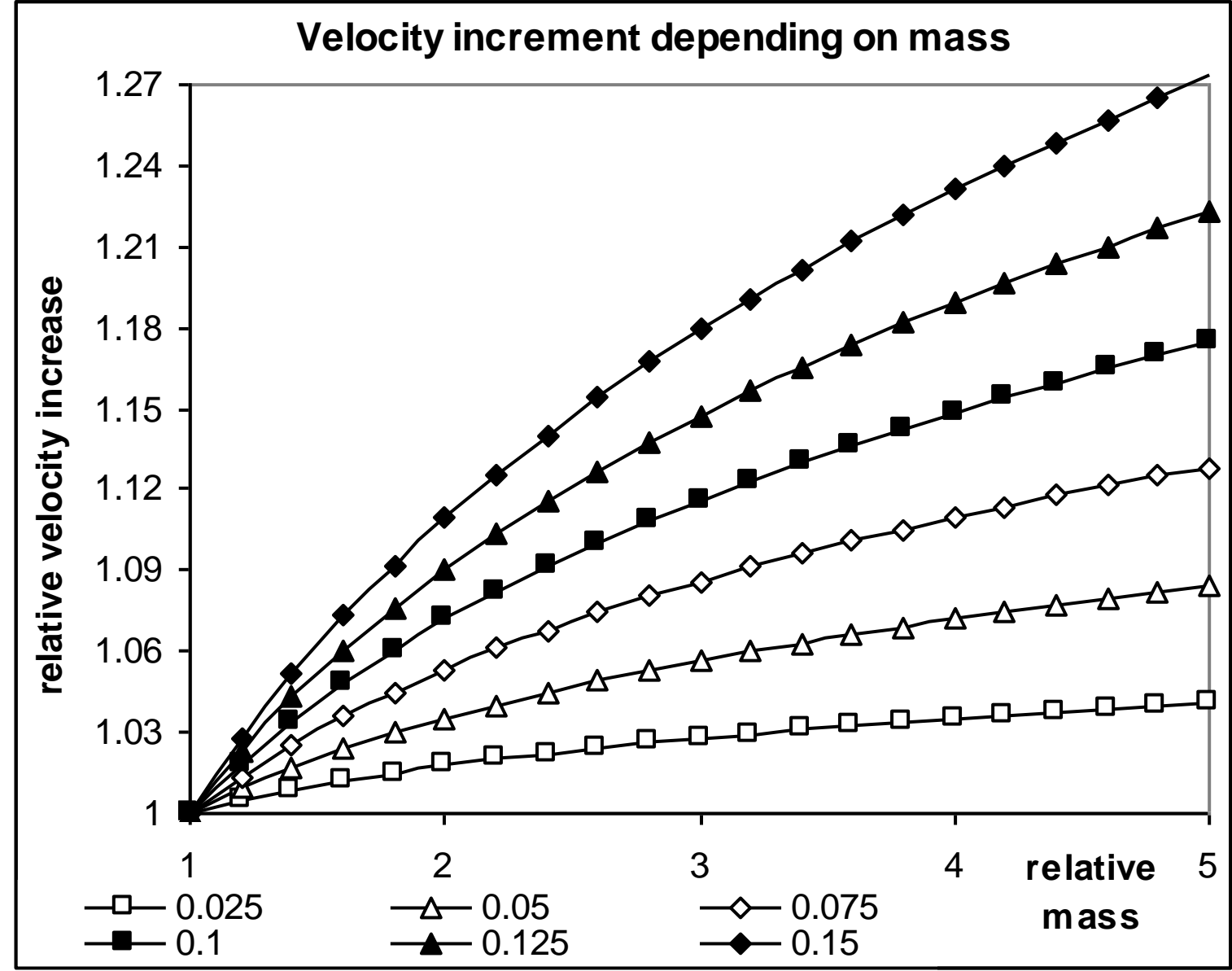

Fig. 3. Relative advantage in velocity of a predator over the prey. Dependence on relative mass and the difference in values of allometric exponents (numbers in the figure legend).

Increase of speed from mouse weighing 44 *g* to animals like caribou weighing 54.5 kilos is from 12.8 $km \cdot h^{-1}$ to 70 $km \cdot h^{-1}$). Using (16), we can find the increment of allometric exponent, due to speed increase, as follows:

$$b - b_t = b_v = \ln(v_x / v_0) / \ln(M_x / M_0) = \ln(70/12.8)/\ln(54.5/0.044) = 0.239 \quad (22)$$

Of course, we can find the speed increase for any value of mass increment; for instance, per certain mass increase corresponding to a well expressed developmental stage.

### 3.3. *Finding interspecific allometric scaling in mammalians*

#### 3.3.1. *Scaling of limb length*

Eq. (18) describes ratio of energies required to support motion from the purely *mechanical* perspective. It includes the ratio of limb lengths. For quadrupedal mammals, it was found in Ref. 32. Depending on the studied group, and fore- or hindlimbs, the allometric exponent noticeably varies. The authors acknowledge that "forelimb slopes range from 0.30 (Rodentia) to 0.42 (Carnivora), while hindlimb slopes range from 0.27 (Rodentia) to 0.42 (Carnivora)". For hindlimbs (which generally consume more energy than forelimbs), the slope is 0.37, for forelimbs, it is 0.4. They say, "Limb length increases disproportionately with body mass via positive allometry 0.40 (length / body mass )".

Our own studies based on data from Ref. 33, 36, and presented in Table 1 and Figs. 4 and 5, produced the value of $0.3655 \pm 0.024$ (it was found as a slope of the regression line calculated for these data presented in logarithmic scale; here and below we use standard errors).

Data in Table 1 include heavier animals than the study in Ref. 32. Weighing these two values by the number of different species (44 and 25), we obtain the value of $0.3875 \pm 0.029$ (here, we combined standard errors as independent values). Data for the animal speed in Table 1 and in all other tables and figures in this work are taken from websites Refs. 37, 38, and checked for validity in various sources.

Table 1. Data for mammalians from Refs. 33-36, 37, 38. used for calculation of allometric exponents for the limb lengths and animals' speed, relative to mass.

| Animal | Mass, *kg* | Limb length, *cm* | Speed, $km \cdot h^{-1}$ |
|---|---|---|---|
| Mouse | 0.044 | 3.5 | 12.8 |
| White rat | 0.21 | 4.9 | 13 |
| Goat | 23 | 42.9 | 17 |
| Horse | 431 | 124 | 64 |
| Elephant | 1542 | 168 | 40 |
| Caribou | 73.5 | 98.8 | 70 |
| Reindeer | 111 | 110 | 80 |
| Musk shrew | 0.036 | 4.5 | 13 |
| Flying squirrel | 0.063 | 7.9 | 24 |
| Chipmunk | 0.092 | 6.7 | 33 |
| Tree shrew | 0.12 | 9.9 | |
| Setifer | 0.12 | 7.4 | |
| Squirrel | 0.21 | 9.8 | 25 |
| Ferret | 0.54 | 13.3 | 25 |
| Dwarf mongoose | 0.58 | 13.1 | 32 |
| Tenrec | 0.68 | 11.9 | |
| Hedgehog | 1.05 | 10.2 | 19 |
| American opossum | 2.7 | 17.8 | 25 |
| Spring hare | 3 | 35.8 | 60 |
| Capuchin | 3.34 | 56 | |
| Suni | 3.5 | 34.4 | |
| Echidna | 3.53 | 14.9 | 30 |
| Cat | 3.9 | 30.1 | 48 |
| Armadillo | 4.07 | 19.4 | 48 |
| Grey wolf | 23.1 | 54 | 75 |

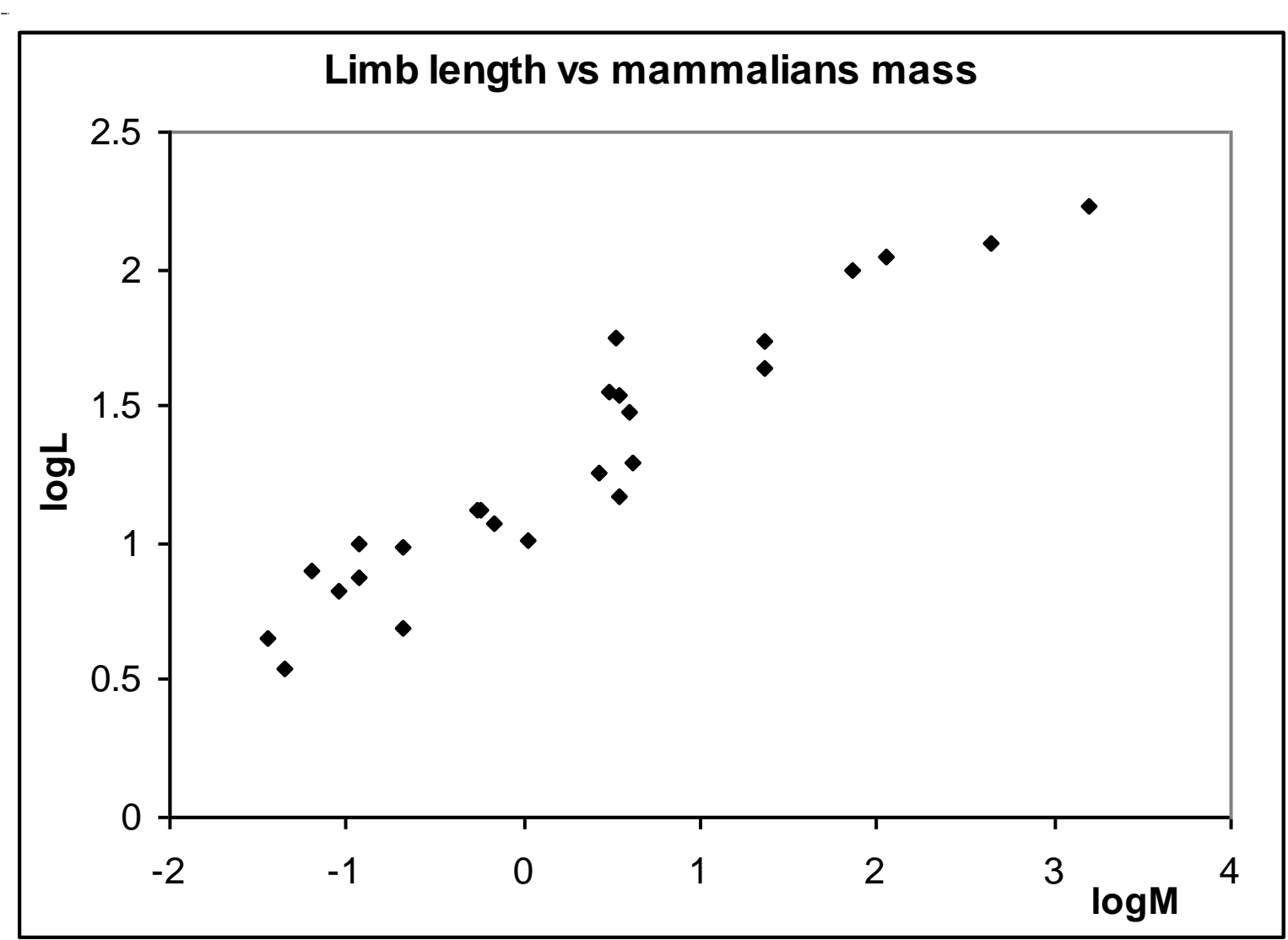

Fig. 4. Data points from Table 1 for finding allometric exponent for the limbs' length increase, versus the animals' mass increase, in logarithmic scale.

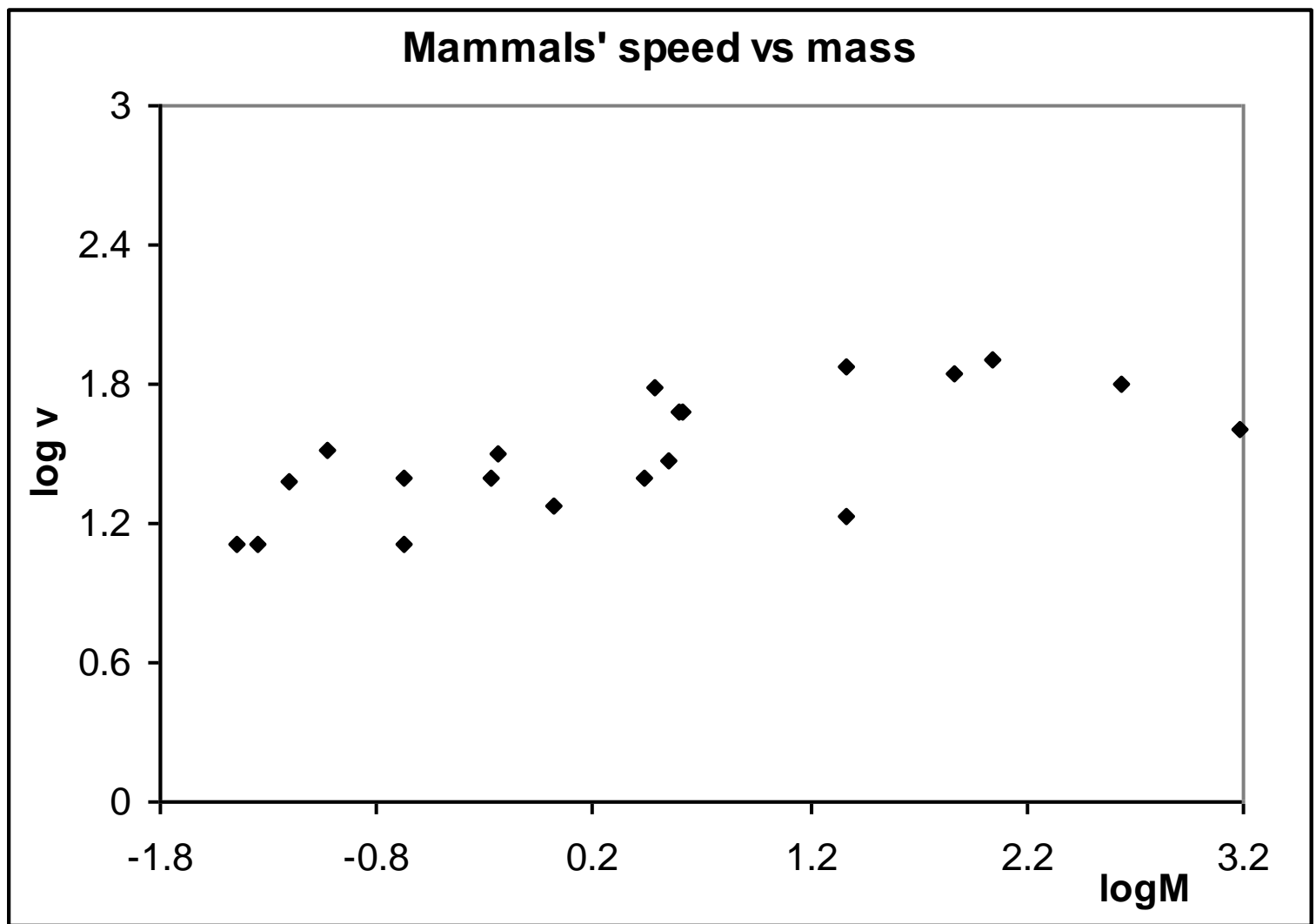

Fig. 5. Data points from Table 1 for the animals speed versus their mass, in logarithmic scale.

### 3.3.2. *Evolutionary speed increase and its conversion to increase of maximal metabolic rate*

When we ascend the food chain and mass of organisms increases, we observe increasing animals' velocity, when they exercise about the *maximal metabolic activity* (chase or flee). Many herbivorous and carnivorous animals coexist in various paired "predator - prey" combinations, so that herbivorous species evolutionarily had to adjust their speeds to predators and vice versa, in order for both populations to not become extinct. Such speed increase in predators, in fact, is a manifestation of the increase of metabolic power of animals when their mass increases. A bigger predator, in order to get a prey, objectively has to have higher speed and power to overcome the prey, and so they were developed and optimized evolutionarily (within the physical, physiological and other constraints). Otherwise, they won't be able to reproduce their population. On the other hand, the predator's speed and power advantage cannot be excessive; otherwise, the prey's population will be quickly destroyed. Keeping this delicate dynamic balance of the food chain is the only way for all organisms to survive together in their concurrent state. Of course, such a balance is not a permanent thing; some species can reduce in quantities, or disappear or excessively reproduce. This will cause the disturbance of the balance, which will be compensated in some way in order to restore the dynamic balance of the food chain in a new way. (Recall the disturbance caused by introduction of rabbits in Australia, for which were no natural predators.)

Of course, the speed is only one possible manifestation of metabolic power. The more critical for the animal's survival the speed is, the more objective measure for the maximal metabolic rate it is. Most species in the range from several tens grams to the order of one hundred kilos rely on speed for their survival. In smaller species, like small lizards, the speed may not be an accurate manifestation of their metabolic power. Such, it was discovered in Ref. 39 that the actual maximum metabolic rate for small lizards is 3.9 times greater than their metabolic capacity required for the maximal running speed on a level surface. Indeed, Fig. 5 shows a noticeable dispersion of velocities for different animals. This is the reflection of the fact that although the maximal metabolic capacity strongly correlates with the speed for many animals, the speed is not the only factor, which animals rely upon for their successful reproduction, but they use all possible means, as it was already discussed in the previous sections.

Big animals like bulls, elephants rely not only on speed, but also on their power and other means, so that their weaponry is more diverse in this regard, and the speed is not necessarily an accurate indicator of their maximal metabolic power. Nonetheless, in situations, when such animals do not need to develop high speed of the whole body, they still have to move their body parts fast to match or exceed the speed of enemies or their preys. It is very similar to the principle of circuit training for athletes, when different groups of muscles are trained at a *maximal* capacity *separately*, and, nonetheless, this accordingly improves the *overall* metabolic capacity[40].

Using data for speed and mass from Table 1, we found the slope of $0.134 \pm 0.033$ for the regression line when data are presented in logarithmic scale. This value is significantly less than the value of 0.21 found in Ref. 31. The difference is explained by inclusion of heavy animals in our dataset, which rely not so much on speed but other means for survival. For another dataset of 13 animals relying more on speed, presented in Table 1B in Appendix B, the allometric exponent is $0.209 \pm 0.032$, which is nearly identical to the value of 0.21 from Ref. 32.

For the dataset 1 of 6 athletic animals from Table 2B in Appendix B, this exponent is equal to $0.255 \pm 0.05$, while for a similar dataset 2 from Table 3 in Appendix B it is $0.392 \pm 0.036$. In other words, its value depends on the choice of animals; to which extent the speed is an accurate manifestation of their metabolic power. For our calculations, we have to use the maximum values for reasonably big datasets, since they are most representative with regard to the speed as a measure of maximal metabolic capacity, that is the value of $0.209 \pm 0.032$ for the general mammalians, and $0.255 \pm 0.05$ for athletic animals.

### 3.3.3. *Calculating MAS for maximal metabolic rate*

Now, we are ready to calculate the interspecific allometric exponent *b* using (18).

$$\frac{E}{e} = \left( \left( \frac{M}{m} \right) \left( \frac{m}{M} \right)^{b_l} \right)^{b_{sk}} \left( \frac{M}{m} \right)^{b_v} = \left( \frac{M}{m} \right)^{(1-b_l)b_{sk}+b_v} \qquad (23)$$

where $b_l$ is the allometric exponent for scaling limbs' length, $b_{sk}$ is the allometric exponent for the skeleton mass, $b_v$ is the allometric exponent for scaling of the animal speed.

So, we found that the allometric exponent for the maximal metabolic rate is

$$b_x = (1 - b_l) b_{sk} + b_v \qquad (24)$$

Substituting the earlier obtained numerical values of allometric exponents into (24), we find the allometric exponent for the general mammalian for a maximal metabolic rate.

$$b_x = (1 - 0.3875) \times 1.08 + 0.2092 = 0.871 \pm 0.042$$

This result very well corresponds to experimentally found value of $b = 0.872 \pm 0.029$ from Ref.41.

The evaluation for athletic animals would be rather subjective, given large difference in the values of allometric exponents depending on the set of animals and smaller datasets. Using the average of 0.3235 for the two obtained values, we find:

$$b_{xa} = (1 - 0.3875) \times 1.08 + 0.255 = 0.985 \pm 0.057$$

In Ref. 41, experiments yielded the value of 0.942 vs ours 0.985, but for a different set of athletic animals. So, there is correspondence between our approximate evaluation and experimental results for athletic animals too. A more accurate study should take into account also scaling of limb lengths for athletic animals, which could be different, and, of course, the dataset of animals for theoretical evaluation should be the same as in experiments. Anaerobic energy producing mechanisms likely contributed to our higher value, since Ref. 41 considered *oxygen* consumption only, while our approach accounts for *all* energy expenditures, and anaerobic energy producing mechanisms contribute the most namely at high levels of physical activity.

There are other experimental studies of allometric scaling for the maximal metabolic rate supporting our results. Refs. 5, 6 quote values of **0.86**, **0.87** and **0.88** for the maximal metabolic rate in mammals, and Ref. 42 also states value of **0.87** for running mammals (from another work). In other words, known *experimental* values are nearly identical as our *theoretical* value of **0.871**, which is definitely a strong argument in favor of validity of our biomechanical model and of the main hypotheses of the study.

### 3.3.4. *Allometric scaling for the basal metabolic rate*

Now, knowing the maximal metabolic rates, we can estimate the basal metabolic rate. For that, we can use the fact that the basal metabolic rate is a stable fraction of the maximal metabolic rate[1,41], usually of about 1/10, although the highly trained athletes may have 1/20 and even 1/30. So, we have to find, how the known fraction of a metabolic power translates into the value of allometric exponent.

Using Taylor series' representation, we can show that for small values of increments of allometric exponent (which is our case), the following formula can be used for approximate estimation.

$$(B_{\max} - B_{mec})k = (aM^{b_{\max}} - aM^{b_{mec}})k = aM^{b_{mec}+(b_{\max}-b_{mec})k} - aM^{b_{mec}} \qquad (25)$$

where $B$ denotes metabolic rate; $M$ is mass, indexes '*max*' and '*mec*' denote the maximal metabolic rate and the 'mechanical' part of the allometric exponent, defined by scaling of limbs and the skeleton mass, equal to $(1-b_l)b_{sk}$; $k$ is the fraction of $b_v$ ($b_v = 0.209$), corresponding to the basal metabolic rate. Mathematical proof of Eq. (25) is presented in Appendix C.

So, in order to obtain the basal metabolic rate $b_b$ for mammals, we should add 1/10 of the average addition of $b_v = 0.209$ due to maximal metabolic activity to the base "mechanical" allometric exponent. This produces the value of

$$b_b = (1-0.3875)\times 1.08 + 0.0209 = 0.682 \pm 0.029 \qquad (26)$$

These results very well correspond to experiments, in particular:

(a) to the value of **0.686** for the basal metabolic rate in Ref. 16;

(b) to the standard metabolic rate (SMR) of **0.678** in Ref. 15 (According to the methodology, SMR should be slightly less than the basal metabolic rate. Indeed, this is what we obtained.)

(c) to value of **0.676** presented in Ref. 42.

Note that the earlier obtained values for the maximal metabolic activity will not change. What happened, we just assigned 1/10 of the increase of the allometric exponent due to maximal physical exercising to the basal metabolic rate, which reflects on animal needs in resources to be alive.

In case of geometric similarity, we obtain the value of $b_b = 0.741 \pm 0.029$, which is very close to Kleiber's original estimation of 0.74.

So, although we accounted for many factors, the theoretically calculated allometric exponents are actually identical to the most accurate available experimental data. Thus, we may conclude that our theory, based on mechanism of optimization of sharing of habitat's resources between species of a food chain by natural selection, adequately explains the origin of interspecific allometric scaling in mammals.

## 4. Interspecific allometric scaling in reptiles, fish and birds

In Appendix D, using the same concept and methodological basis, allometric exponents for reptiles, fish and birds were found. For fish, reptiles and birds we had less data. The reason is that, however surprisingly this may be, most data we could use for the purposes of the study, were obtained in earlier days, and since then such experiments, to the best of our knowledge and efforts to find newer material, were not conducted. So, it is rather not negligence or lack of desire to find the data, but the scarcity of data themselves. Even though, we still were able to confirm that the same mechanism of optimal sharing of resources between the species of a food chain, guided by natural selection, indeed, defines the interspecific allometric scaling of these classes of living organisms too. The obtained results correspond to available experimental data well[1,3,15,42]. There is some diversion (in the range of (0.07-0.11)) of calculated value of allometric exponent for the maximal metabolic rate of birds. In our view, this effect is explained by specific features of birds' motion, such as moving in a 3-D space, and the need to spend energy to counteract gravitational force during the flight. Detailed consideration of this issue is in Appendix D.3.

Summary of numerical results for all considered classes of animals is presented in Table 2.

## 5. Discussion

We presented a theoretical study of origin of allometric scaling in mammals, which is well supported by accurate correspondence of calculated theoretical values of allometric exponents to experimental data. This correspondence is a strong argument in favor of validity of the main discovery of the study:

> *Metabolic allometric scaling of multicellular organisms is a consequence of natural selection, which optimizes sharing of habitat's common resources between species of a food chain populating this habitat. The optimization is directed towards supporting stability of a food chain, in order to each species could continually reproduce, while at the same time being a sharable resource for the rest of the food chain. This arrangement preserves a delicate dynamic balance of distribution of habitat's resources allowing perseverance in time and space of species composing* *the food chain, while avoiding overexploitation of common resources by one or several species to the detriment and even possible extinction of other species. Metabolic allometric exponent is a quantitative reflection of how different species within a food chain optimize, through selection, sharing of common energy resources of a given habitat. The process of selection and optimization is framed by the boundaries defined by physical, environmental, physiological, organismal constraints at any given moment.*

The importance of the aforementioned balance is that it secures that all animals in the balanced food chain are able to obtain sufficient amount of food for the reproduction of their populations, without jeopardizing reproduction of populations of other species of the food chain.

### 5.1. *Summary of main results*

Besides the presented main result of the study, another important development is that de facto we created a *framework* for similar studies of other taxa.

We found that the total interspecific allometric exponent is the result of action of several factors, when the mass of organisms increases:
(a) Scaling of limb masses (tails for fish);
(b) Distribution of inertial masses between moving limbs;
(c) Scaling of limb lengths;
(d) Scaling of skeleton masses;
(e) Scaling of the maximal metabolic power;
(f) Fractions of the maximal metabolic power corresponding to basal and other specific metabolic rates.

These factors are discussed in detail in Appendix E.

The results of general importance, discovered in the study, are as follows.

(a) Interspecific allometric scaling is a quantitative reflection on how common resources of a habitat are shared between the species composing a food chain. This distribution of common resources is optimized by natural selection to provide stability of a food chain. Living organisms adjust their metabolism in a way that preserves continuity of a food chain through reproduction of sufficient quantities of each species. Animals, whose nutrient acquisition depends on motion, perform movements with the speed, force and duration adequate to acquire sufficient amount of food for a successful reproduction, but at the same time without destroying populations of their preys, thus supporting a dynamic balanced state of the food chain. The direct "predator - prey" relationships in the food chain are most visible ones. However, organisms in the food chain also relate to each other through numerous feedback loops, through common nutritional environment and other common environmental parameters. Thus, the optimized distribution of resources between the species of a food chain should not be confined to direct "predator - prey" relationships only. Such an arrangement is the principle, which guides how organisms within the food chain organize and evolutionarily develop.

(b) Allometric scaling of energy requirements due to mechanical constraints imposed on animals' propelling extremities, the whole body and associated body parts propagate through the entire organism, because of the primary importance of motion characteristics for successful reproduction (for animals, whose nutrient acquisition depends on motion).

(c) Mechanical characteristics of propelling extremities adjust to evolutionarily and *functionally* optimal motion characteristics of animals (but not mechanically optimal); in many instances, these characteristics relate to animals' speed as a primary factor supporting their existence and successful reproduction.

(d) Evolutionarily, the metabolic power increases with the increase of mass (in particular, expressed as an increase of animals' velocities in a certain range of sizes), since a bigger animal has to overcome its prey (if it is a predator) or be able to protect itself from predators (like big herbivores). Such an objective and measureable increase of metabolic power is propagated through the food chain in all life domains we studied (including unicellular organisms). Greater metabolic power can be manifested by different adaptation means, of which the speed is an important, but not the only one. On the other hand, the need in a balanced state of a food chain caps the increase of metabolic power of bigger animals. Of course, life developed many forms. However, the size increase was and continues to be a backbone of the evolutionary process, from which other developmental branches followed. (Imagine evolution without the size increase!)

### 5.2. *Overview of obtained allometric exponents*

The summary of calculated allometric exponents versus experimental data available in the literature is presented in Table 2.

Table 2. Summary of obtained allometric exponents versus experimental and other studies. MMR - maximal metabolic rate (MR); BMR - basal MR; SMR - standard MR; FMR - field MR.

| Taxa | MMR Model | MMR Experimental | BMR Model | BMR Experimental |
|---|---|---|---|---|
| Mammals | **0.871** | 0.872 [41]; (0.86-0.88) [5,6,42] | **0.682** | 0.686[16]; 0.678[16]; 0.676[42]; 0.678[15] (SMR) |
| Athletic mammals | **0.985** | 0.942 [41] | No data | No data |
| Reptiles [a] | **0.92** | 0.889 [45], 0.89-0.97 [6] | **0.767** | 0.768 [15]; 0.76[42] (FMR) |
| Fish | **0.978** | 0.97[50], 0.974 [6] | No data | No data |
| Birds | **0.771** | 0.84, 0.88 [6] | **0.651-0.657** | 0.652[13]; (0.69, 0.681, 0.667)[6]; 0.644[15] (SMR) |

[a] For the estimated value of allometric exponent for the skeleton mass of 0.898.

Theoretical results were obtained by estimating energy expenditures for organisms with different mass, through association of species' velocities with metabolic energy.

Calculated allometric exponents correspond well to experimentally found values. According to proposed in subsection 2.1.3 methodology of verification of our hypotheses and results, this makes a strong case that, in the context of our study, the metabolic allometric scaling, indeed, is shaped by mechanism of natural selection, which optimizes distribution of common resources of a habitat between species composing its food chain. This optimization is performed in a way, securing stability of a food chain by sustainable reproduction of composing it species, which acquire optimal for that purpose share of common resources. The logic of this verification approach is shown in Fig. 6.

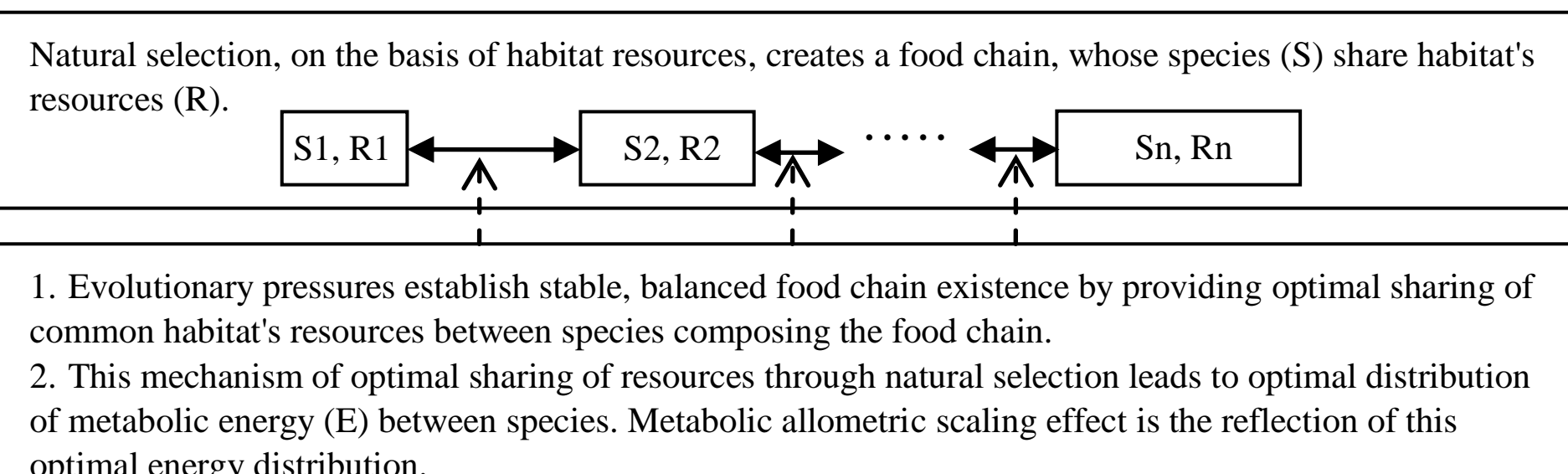

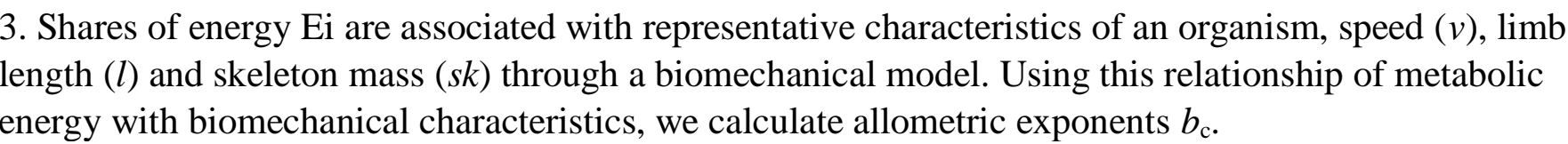

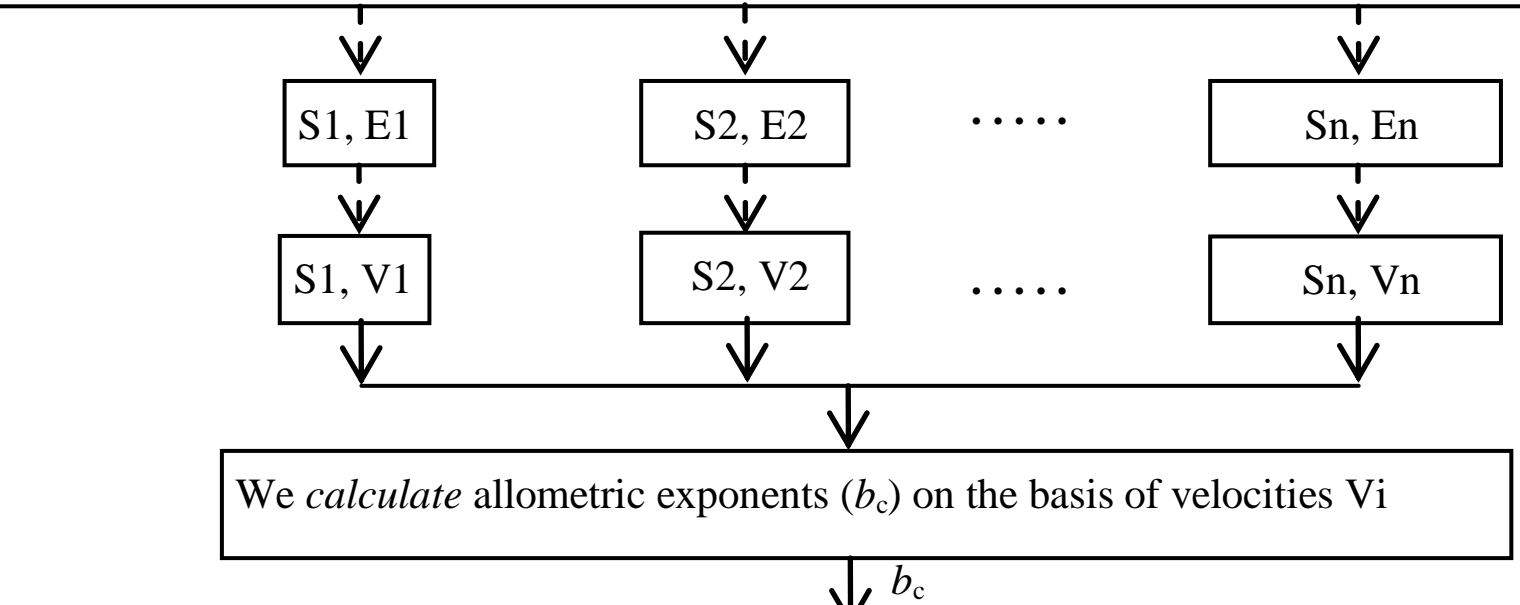

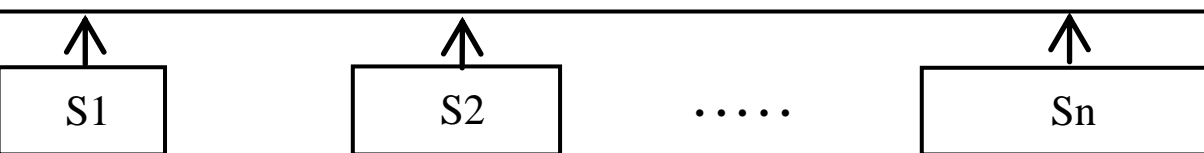

Fig. 6. Scheme of validation of original hypothesis that metabolic allometric scaling, and accordingly the allometric exponent, are defined by natural selection, which optimizes distribution of common resources of a habitat between species composing its food chain.

## 6. Conclusion

One of the results of this study was discovery of principle of evolutionary continuity of a food chain and its dynamic balance as a condition of its existence and of each organism there. This is not a world that knows no restrictions and rules. It is as much destructive as a constructive one; this is a balanced world, in which *measure* is the norm, the rule of the game, but not the exception. These rules will prevail over the long term, despite deviations. There are many important things in this arrangement, which humans could learn for their benefit (but they will not).

The presented concept and methods can be used for other groups of living organisms, in different strata; not only for classes of animals, but also for smaller or larger stratifications created on the basis of different criteria.

In our opinion, this study opens new promising areas. Effectively, the proposed framework for scientific studies and practical applications can be used in many areas, related to biology and medicine.

One of the important consequences of the study was the recognition that the seemingly particular problem of interspecific allometric scaling all of a sudden became tightly connected to physiological and evolutionary principles of life origin, development and organization at the level of a food chain.

The other consequence of this study is that it shifts the focus of the present biological paradigm from particular physiological and biochemical mechanisms, as determinants of organic life, and of interspecific allometric scaling in particular, to better understanding that all these mechanisms are supplementary to providing a great range of adaptive flexibility, which allows organisms to adapt to a wide range of environments and mutual coexistence. If we think for a moment, it could not be otherwise, since all these mechanisms were developed during evolution, and belong to organisms, which manage to survive in harsh environments, and their adaptation capabilities allow to adapt to the *present* environmental conditions. (If not, they would not be with us anymore.) When environmental conditions will change, organisms, through natural selection and optimization, will develop new adequate adaptation mechanisms, if the old ones are insufficient, or they will perish.

This continuous adaptation is guided by certain rules, determinants of high level, which together provide evolution of organisms within the food chain, whose continuity and dynamic balance is a condition of survival of every one and all organisms together. This balance by its nature is a dynamic one; links can disappear, new links can emerge, but the food chain will restore its continuity and establish a new balance, because these are inherent properties of organization of the organic world as a whole, the fundamental principles its existence is based upon.

**Acknowledgements**

The author thanks Dr. A. Y. Shestopaloff for fruitful discussions and help with editing, and Dr. K. Y. Shestopaloff for the help with data processing, providing the literature, for the review and valuable suggestions, which helped to enhance paper's content, structure and clarity. Editor's invitation to submit the paper to BRL is very much appreciated, as well as efficient handling of the review process. The author is thankful to Reviewers, whose comments helped to better expose article's ideas and approaches, and to make the content of the paper accessible to a wider audience.

## Appendix A. Increase of metabolic capacity with the increase of skeletal mass

Let us imagine an extreme scenario. Suppose that we found three people with successively increasing weight of 60, 70 and 80 kg, and with the same relative skeleton mass of 30% (accordingly 18, 21 and 24 *kg*). All of them have the same maximal oxygen consumption per kilo per minute. Let us simulate increase of the relative skeleton mass assuming that now participants have a relative skeleton mass accordingly 30%, 40%, 50%. Then, the second participant should put a backpack with 7 *kg* (70x0.4 - 21 = 7), and the third one 16 *kg* (80x0.5 - 24 = 16). Let them carry bricks, weighing 30% of their body without skeleton mass (that is 12.6, 14.7, 16.8 *kg*) upstairs in a high-rise construction, for several months. Then, the overall maximal oxygen consumption is tested again. Although this parameter is considered as a conservative one, one will find a substantial increase with each simulated skeletal mass increment, even if we account for some possible weight gain. This fact was proved by studies of endurance physical training, with assessments of outcomes for several parameters, including the oxygen consumption[40, 43].

Thus, adding the passive weight gives muscles and other constituents an additional metabolic training, which supersedes some possible weight gain, and often occurs even without such a gain.

## Appendix B. Data for mammals

Table 1B. Animal mass and speed used for calculation of allometric exponent for speed relative to mass.

| Animal | Mass, *kg* | Speed, $km \cdot h^{-1}$ |
|---|---|---|
| Mouse | 0.023 | 12 |
| Rat | 0.4 | 13 |
| Rabbit | 3 | 48 |
| Cat | 4.5 | 45 |
| Gray fox | 7.00 | 48 |
| Coyote | 13 | 65 |
| African Wild Dog | 27 | 72 |
| Greyhound | 32 | 63.5 |
| Pronghorn | 46 | 88 |
| Lion | 175 | 80 |
| Elk | 320 | 72 |
| Horse | 430 | 64 |

| Red kangaroo | 55 | 70 |
|---|---|---|

The allometric exponent calculated from data in Table 1B for scaling of velocity relative to mass is equal to $0.209 \pm 0.032$.

Table 2B. Mass and speed of athletic animals. Dataset 1.

| Animal | Mass, *kg* | Speed, $km \cdot h^{-1}$ |
|---|---|---|
| Cat | 4.5 | 45 |
| Gray_fox | 7.00 | 48 |
| Coyote | 13 | 65 |
| African_Wild_Dog | 27 | 72 |
| Greyhound | 32 | 63.5 |
| Pronghorn | 46 | 88 |

Allometric exponent for the speed relative to mass in athletic animals from Table 2B is equal $0.255 \pm 0.05$.

Table 3B. Mass and speed of athletic animals. Dataset 2.

| Animal | Mass, kg | Speed, $km \cdot h^{-1}$ |
|---|---|---|
| Gray_fox | 7 | 48 |
| Coyote | 13 | 65 |
| African_Wild_Dog | 23 | 72 |
| Pronghorn | 46 | 98 |
| Cheetah | 52 | 112 |

Allometric exponent for the speed relative to mass in athletic animals from Table 3B is equal to $0.392 \pm 0.036$.

**Appendix C. Mathematical proof of Eq. (25) (translation of fraction of metabolic power into a fraction of allometric exponent)**

Eq. (25) means that the fraction of metabolic rate *k* for small increments of allometric exponent can be assumed as equal to a fraction of increment *p* of the allometric exponent. It can be proved as follows.

$$(B_{\max} - B_{mec})k = (aM^{b_{\max}} - aM^{b_{mec}})k = kaM^{b_{mec}}(M^{(b_{\max}-b_{mec})} - 1) \approx$$

$$kaM^{b_{mec}}(1 + (b_{\max} - b_{mec})\ln M - 1) = kaM^{b_{mec}}((b_{\max} - b_{mec})\ln M) = aM^{b_{mec}} \ln M^{k(b_{\max}-b_{mec})} \approx$$

$$aM^{b_{mec}} \ln(1 + (M^{k(b_{\max}-b_{mec})} - 1)) = aM^{b_{mec}}(M^{k(b_{\max}-b_{mec})} - 1) = aM^{b_{mec}+k(b_{\max}-b_{mec})} - aM^{b_{mec}}$$

Here, we used the Taylor series' expansions for the terms $M^{(b_{\max}-b_{mec})}$ and $\ln M^{k(b_{\max}-b_{mec})}$. The accuracy of approximation is largely defined by the third term of the Taylor series for $M^{(b_{max}-b_{mec})}$ and by the value of $\ln M$ . When $M$=2, for our numerical data, the value of this term is about $((b_{max}-b_{mec})\ln M)^2/2 \approx 0.006$, which is a negligible error.

Of course, the exact fraction $p$ of increment of allometric exponent (above the value of the 'mechanical' allometric exponent) corresponding to the basal metabolic rate, can be accurately found by solving numerically the following equation.

$$(aM^{b_{max}} - aM^{b_{mec}})k = aM^{b_{mec}+(b_{max}-b_{mec})p} - aM^{b_{mec}}$$

It can be transformed to a more convenient form as follows.

$$kM^{b_{max}-b_{mec}} - k + 1 = aM^{(b_{max}-b_{mec})p}$$

Solution of this equation for $k$=0.1, $M$=2 and $b_{max}-b_{mec}=0.16$ is $p=0.105$ (versus the compared value of $k=0.1$); for $M$=4 the solution is $p=0.11$, so that, indeed, given the fact that we apply this fraction to values of the order of 0.1, that will be a negligible error. Thus, for our estimation purposes, we can use Eq. (25).

On the other hand, since the value of $k$ varies depending on the species, such variability apparently provides variability of allometric exponents for the basal metabolic rates in different organisms. For instance, organisms living in cold climates might have higher basal metabolic rates, and accordingly greater values of $k$. Overall, this is the issue to be studied further. For now, we want to know if our theory produces values of allometric exponent for the basal metabolic rate corresponding to experimental measurements or not.

## Appendix D. Interspecific metabolic allometric scaling in reptiles, fish and birds

### D.1. *Allometric scaling in reptiles*

For reptiles, we used a limited set of data for 12 specimen: from Ref. 33 in order to find scaling of limb lengths relative to mass, and from Refs. 33, 39 to find the scaling of speed relative to mass.

Table 1D. Mass and limb lengths for reptiles. Data from Ref. 33

| Reptiles | Mass, *kg* | Limb length, *cm* |
|---|---|---|
| Lizard | 0.008 | 3.3 |
| Green_lizard | 0.026 | 3.7 |
| N-desert_iguana | 0.051 | 4.8 |
| Australian_skink | 0.47 | 6 |
| Lizard | 0.006 | 3.2 |
| Lizard | 0.01 | 3.4 |

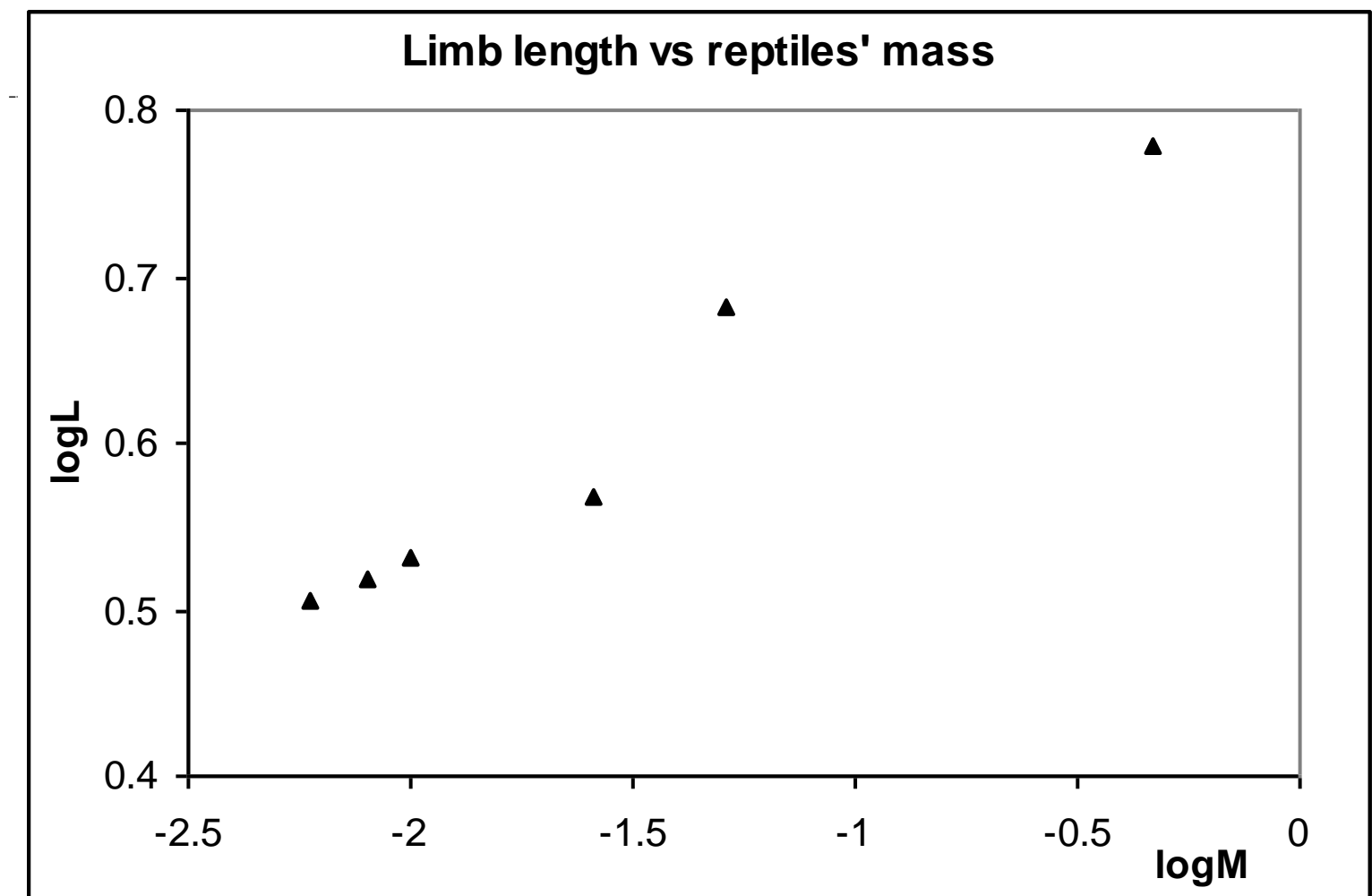

Fig. 1D. Scaling of limb length versus mass in reptiles. Data from Table 1D.

Table 2D. Mass and adjusted speed for reptiles. Data from Refs. 38, 39

| Reptiles | Mass, *kg* | Speed, $km \cdot h^{-1}$ |
|---|---|---|
| Frilled lizard | 0.9 | 20 |
| Tuatara | 0.75 | 24 |
| Caiman | 250 | 48 |
| Monitor lizard | 160 | 45 |
| C. variegatus | 0.052 | 14 |
| E. skiltoniaus | 0.042 | 10.68 |

The data show well expressed trends with very reasonable diversion from trending lines in both instances. Speed for lizards *C. variegates* and *E. skiltoniaus* was increased by 3.9 times according to estimations of the real metabolic capacity done in Ref. 39.

Scaling of the skeleton mass relative to the total lizard mass was studied in Ref. 44. Skeletal mass varied from 0.15 to 972.75 *g*. The authors conclude "in lizards, skeletal mass scales … negatively allometrically with body mass". The estimated value for the slope was 0.716. On the other hand, the authors found that "Body mass and SVL ("snout-vent length") were highly correlated, and the slope of the RMA regression was not different from the slope predicted for isometry. Similarly, skeletal mass and SVL were highly correlated, and again the slope of the RMA regression was not significantly different from the prediction for isometry."

In this quotation, one can see inconsistency. If SVL scales about isometrically with the body mass, and the skeletal mass scales about isometrically with SVL, then the scaling of the skeletal mass to the body mass should be also close to isometrical, and cannot be as negative, as the authors reported. Note that the negative allometry, claimed by the authors, was deduced on a small amount of data (the authors acknowledge, "Skeletal mass scaled with negative allometry against body mass for this small subset"). Thus, the inference about close to isometric scaling of limbs' mass looks as a more reliable result, than negative allometry reported on a small amount of data.

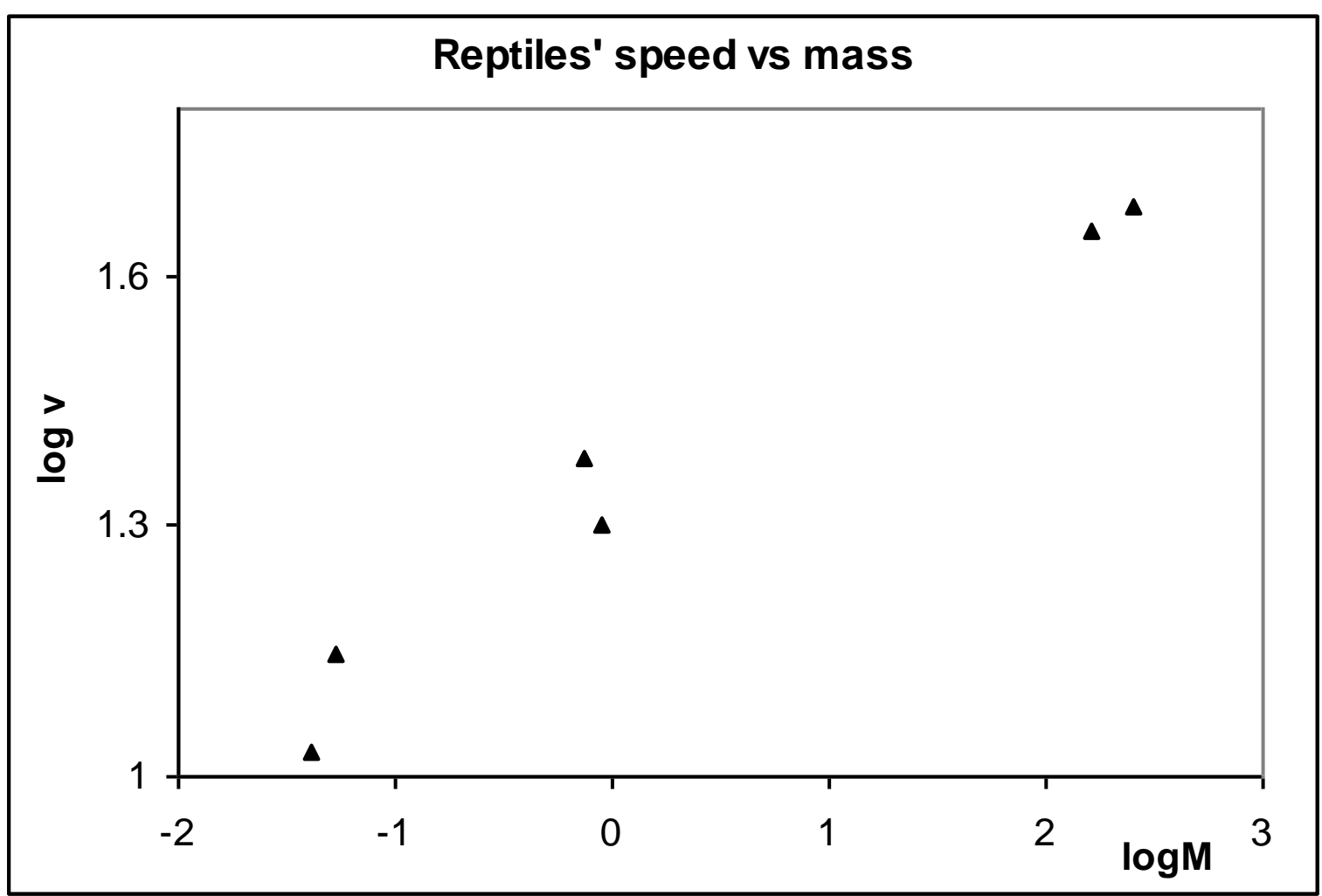

Fig. 2D. Reptiles' speed versus mass in logarithmic scale. Data from Table 2D.

Regression analysis of available data produced the following. Allometric scaling for limbs is $b_l = 0.151 \pm 0.016$, for speed $b_v = 0.157 \pm 0.017$. For reptiles, we assume that we do know neither the allometric exponent $b_{sk}$ for the skeletal mass (except that it might be slightly negative), nor the fraction *k* (corresponding to basal metabolic rate) of the increment $b_v$, corresponding to maximal metabolic capacity. However, we can find both values solving the system of two equations, derived from Eq. (24) in the main text. In Ref. 45, the authors obtained the allometric exponent of 0.889 for active reptiles, although there is no guarantee that this was the maximal possible metabolic activity. On the other hand, In Ref. 15 authors obtained the value of 0.768 for the standard metabolic rate in reptiles (which is close to basal metabolic rate). So, we can write the system of equations as follows.

$$(1-b_l)b_{sk} + b_v = b_x = 0.889$$

$$(1-b_l)b_{sk} + kb_v = b_b = 0.768 \qquad \text{(D.1)}$$

Substituting the found values of $b_l = 0.151$ and $b_v = 0.157$, and solving the system of equations, we find $b_{sk} = 0.862$, $k = 0.23$. These numbers are doubtful to be true. In mammals, the value of *k* is about 1/10 and less, so that it is unlikely that in ectotherms, some of which can live without food for weeks and even months, the value of *k* is as high. The value $b_{sk} = 0.862$ seems as a little bit low too.

Experiments show that a greater value of allometric exponent for the maximal metabolic rate in reptiles is possible. In review Ref. 6, the author mentions "squamate reptiles exhibit a greater range of scaling exponents (0.27 - 1.26)". Since $k < 1$, then it follows from the system of equations (D.1) that for the specimen, which we considered, $b_x - b_b < b_v$, that is $b_x < 0.925$. For instance, if $b_x = 0.92$, then $k$=0.032, which seem as possible numbers for reptiles. Indeed, for nine specimen of varanid lizards the allometric exponents were found to be 0.89 and 0.97 for 25 and 35 $^0$C accordingly[6], so that the obtained numbers are realistic ones. Then, in this case, we have the following dataset: for the maximal metabolic rate $b_x = 0.92$, for the basal metabolic rate $b_b = 0.767$, and the allometric exponent for the skeletal mass is $b_{sk} = 0.898$. The unintended surprise for this scenario is that this value of allometric exponent $0.767$ very well matches the value of allometric exponent of 0.768 for the standard metabolic rate in reptiles obtained in Ref. 15, which we consider as a reliable source.

Thus, although for reptiles we did not obtain the allometric exponents for the maximal and basal metabolic rates directly, like for mammals, we showed that our results agree with available experimental data. Besides, even though we have had two unknown parameters, the proposed method still allowed attaining useful results, predicting important characteristics, and providing cross-verification of obtained values. It's probably even more important that the discovered metabolic integrity of a food chain, created through natural selection and optimization, and developed on its basis methods provide clear directions for future studies of metabolism in reptiles, setting priority for:
(a) Finding the scaling of skeletal mass relative to the total mass;
(b) Finding the fraction of the maximal metabolic rate corresponding to the basal metabolic rate, which then can be compared to values we obtained.

### D.2. *Allometric scaling in fish*

For fish, the speed is an objective indicator of the metabolic power, since the water resistance in the range of velocities the overwhelming majority of fish swim in can be considered proportional to speed. This assumption is further enforced by the physical effect of mutual compensation of circular water turbulences, vertexes, arising on opposite sides of a fish body. This happens due to movements of a fish tail, which swap these turbulences, so that the one, originated on the right side, is moved to the left side, and vice versa. With such an arrangement, the turbulence vertexes meet behind the fish by flows moving in the opposite directions, thus canceling each other. (A popular video about this interesting effect is at https://www.youtube.com/watch?v=wYDh5d9pfu8.) It might be one of the main reasons why fish can swim so economically and tirelessly, unlike, say, humans. In addition to this effect, at extreme velocities, fish and sea animals like dolphins can further reduce turbulence by forming sort of wrinkles on the skin, thus extending the range of speeds where the water current around fish bodies still resembles laminar flows. Such assumption also agrees with results of studies from Ref. 46, reporting a narrow range of Strouhal number, characterizing high energy efficiency of fish swimming.

The main contours of bodies of fast swimming fishes were evolutionarily shaped more or less similarly, much due to the need to adapt to hydraulic resistance during motion and specific way of motion through water. Although the relationships between the body length and volume are different for different fishes, the same species scale in size in a way close to geometric similarity. Such, in Ref. 47, the author found that the length relates to mass (regardless of the size) at a power of 2.8 for dace, 3 for trout and 3.2 for goldfish. Our own study based on photos from books and the Internet showed that in fast fishes of different sizes, from centimeters to several tens of centimeters, the proportions between lengths of tails and bodies do not vary significantly. This is apparently the consequence of the uniform body optimization to overcome the hydraulic resistance by the mentioned earlier swapping of circular turbulences by movements of tail, and also to adjust the overall body shape to minimize water resistance. So, for our estimation purposes, we assumed the geometric similarity. However, the influence of scaling of tail length and surface with the increase of size requires further studies. Besides, fish use body movements for the propulsion too.

For fish, the equivalent of limb length, which we used in our formulas for mammals, is the length of tail, which is the main propeller in fish. The equivalent of the length of strides in mammals for fish is the tail beat frequency. It was studied in Ref. 48. The results showed that the speed is directly proportional to the beat frequency. This may be not so surprising, but nonetheless a remarkable result for our purposes, which means that the "mechanical" part of the allometric exponent in fish can be described by the same formulas, which were derived for mammals (except for the vertical oscillations, which we do not have in fish). Similar considerations can be applied to fins, whose length we assumed to increase proportionally to the length of fish. (Even if there are some diversions from this assumption, they are not of importance for our purposes, since the tail is the main fish propeller at high speeds, when fish manifests the highest metabolic power.)

We used data from Refs. 47, 48 (total 33 fishes, divided into two datasets of 14 and 19), in order to find the allometric exponent related to speed increase. (Few measurements were discarded when fish obviously swam below maximum speed.) Data in tabular and graphical forms are presented below.

Table 3D. Mass and speed of fish. Data from Ref. 47.

| Name | Mass, $g$ | Speed, $cm \cdot s^{-1}$ |
|---|---|---|
| Bleak | 1 | 50 |
| Carp | 5 | 52 |
| Carp | 6 | 59 |
| Sea trout | 34.1 | 92 |
| Mackerel | 25.2 | 81 |
| Twaite shad | 29.7 | 75 |
| Perch | 18.4 | 66 |
| Meagre | 29.5 | 113 |
| Bib or Pout | 16.5 | 55 |
| Grey mullet | 26 | 61 |
| Rudd | 18.8 | 114 |
| Hake | 23.7 | 79 |
| Tuna | 27240 | 1959 |
| Southern ground shark | 9534 | 405 |

Table 4D. Mass and speed of fish. Data are from Ref. 48.

| Name | Length, *cm* | Value $\propto M$ | Speed, $cm \cdot s^{-1}$ |
|---|---|---|---|
| dace | 10 | 631.0 | 133 |
| | 10.4 | 704.2 | 115 |
| | 14.5 | 1785.8 | 160 |
| | 15.2 | 2037.8 | 180 |
| | 16.7 | 2652.2 | 200 |
| | 20 | 4394.2 | 225 |
| | 21.4 | 5310.8 | 240 |
| trout | 10.3 | 1092.7 | 106 |
| | 15 | 3375.0 | 175 |
| | 20 | 8000.0 | 175 |
| | 28 | 21952.0 | 270 |
| goldfish | 6.7 | 440.0 | 75 |
| | 9.2 | 1213.7 | 104 |
| | 9.7 | 1437.7 | 114 |
| | 11.8 | 2691.7 | 104 |
| | 13.5 | 4140.6 | 129 |
| | 14.8 | 5557.0 | 140 |
| | 16 | 7131.6 | 188 |
| | 21.3 | 17816.2 | 200 |

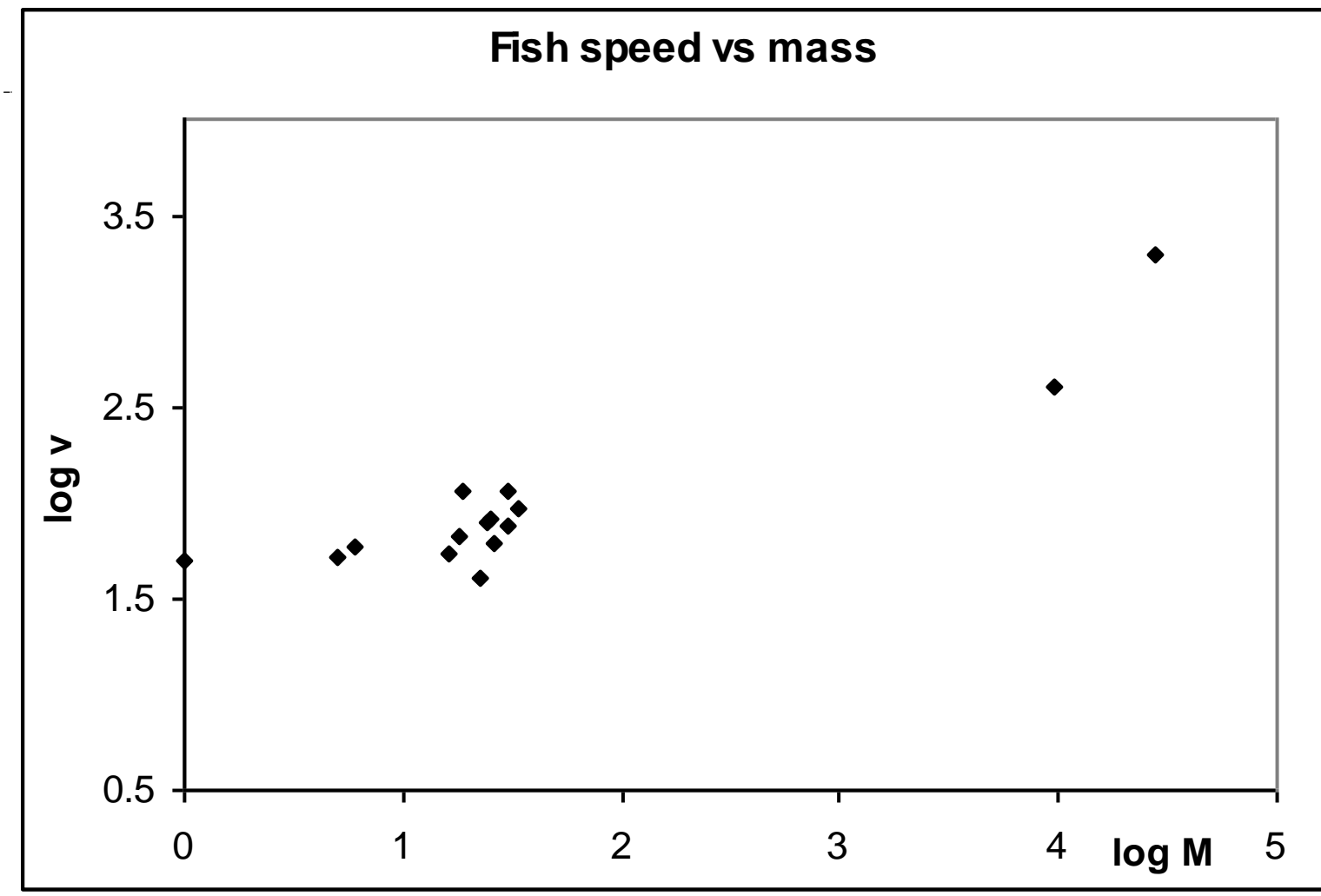

Fig. 3D. Fish speed versus mass for data from Ref. 48, in logarithmic scale.

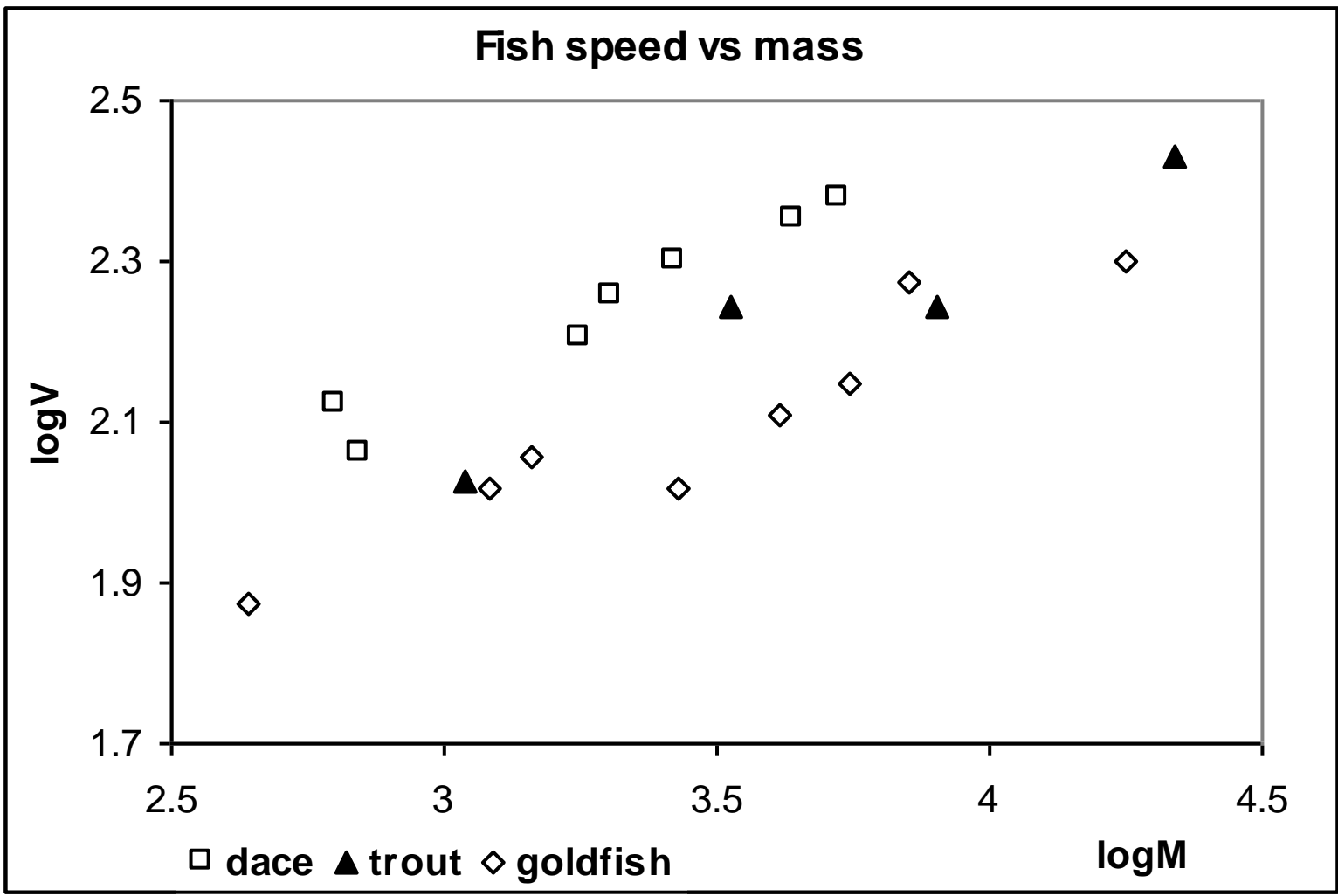

Fig. 4D. Fish speed versus mass for data from Ref. 47, in logarithmic scale, for dace, trout and goldfish.

The graph for the 14 fish dataset shows that the suitability of data in this case is rather questionable, although the trend is clearly expressed. As the author of Ref. 48 mentioned, it is difficult to figure out, if a fish swims at a maximum possible speed, and this factor might be the cause of observed dispersion of data.

The value proportional to mass for the 19 fish dataset was calculated as the fish length at a power of 2.8 for dace, 3 for trout and 3.2 for goldfish, according to Ref. 47.

Allometric scaling for speed for the first dataset is $b_v = 0.343 \pm 0.038$; for the second dataset the value of $b_v = 0.291 \pm 0.076$ (0.32 for dace, 0.288 for trout and 0.244 for goldfish). In the last case, the average slope was found as the average of slopes for three fishes. This was done because each fish has a somewhat different shape, which results in different water resistance. As a consequence, the intercepts of regression lines are different for all three fishes. Thus, finding the average slope as an average of three slopes is more appropriate in this case, than calculating a single slope for all fishes.

The scaling of skeleton mass in fish with increase of mass, according to Ref. 49, is equal to 1.03. Now, we can use Eq. (24) and calculate the allometric exponent for the maximal metabolic rate for the second dataset as follows.

$$b_x = (1 - 0.3333) \times 1.03 + 0.291 = 0.978 \pm 0.076 \qquad \text{(D.2)}$$

For the first dataset, this value is $b_x = 1.03 \pm 0.038$. For small fish, with the exclusion of the 1 g marginal "champion", the result is $b_x = 0.95 \pm 0.11$. So, according to our calculations, the possible range of allometric exponents for fish is $0.687 < b \leq 1.03$ ($(1 - 0.3333) \times 1.03 = 0.687$).

These results comply with experimentally found value of allometric exponent of 0.97 for the maximal metabolic rate, and 0.78 at rest, for a sockeye salmon[50]. In review Ref. 6, the value of 0.974 for the rainbow trout is given; in the same review, the reprinted Fig. 4A shows regression lines for different teleostic fishes, whose allometric slopes range from about 2/3 to slightly over 1. The values of $0.87 \pm 0.078$, and 0.793 as the mean scaling exponent, are quoted too. In Ref. 15, the authors obtained the value of 0.879 for the standard metabolic rate in fish, normalized to 38 $^0$C and 20 $^0$C.

All these numbers are within the range, which we obtained. Of course, it would be very useful for verification to find a basal metabolic rate, but, unlike in case of mammals, we have no estimations, which fraction of the maximal metabolic rate it constitutes, and it is likely that this fraction is different in different fishes. If we assume that the standard metabolic rate is equal to the basal metabolic rate, then, using Eq. (24) and the value of 0.879 from Ref. 15, we can estimate the required increment for the basal metabolic rate above the base value as follows: $0.879 - (1 - 0.3333) \times 1.03 = 0.192$, or about 66% of the maximal metabolic rate increase, which is probably high. For the value of mean allometric exponent of 0.793 from review Ref. 6, this will give 36%, which is - maybe - still high. On the other hand, given the specific environment fish live in, and the highly economical way of moving in water, maybe this number is not unreasonable.

Is such high basal metabolic rate a specific feature of fish, or some other factors are involved, like scaling of tails' lengths and surfaces? We have no answer. If the tail increases slower than the whole body, then the "mechanical" constituent of the allometric exponent for fish will be greater, which will accordingly reduce the fraction for the basal metabolic rate. (Photos of tuna, indeed, show that the relative length of its tail is less than in smaller fish.) So, further studies are needed, which can go two ways: The best one would be to know how the tail lengths scale with the mass increase. Another approach would be to find the fraction corresponding to the basal metabolic rate, and then, using Eq. (24), calculate the scaling exponent for tails.

In any case, the outcome of the fish study is that the allometric exponent for the maximal metabolic rate, which we found, is in a good agreement with published experimental data, while finding the allometric exponent for the basal metabolic rate requires further studies.

### D.3. *Allometric scaling in birds*

For birds, we should find the scaling of wing span depending on the birds' mass, scaling of wing mass relative to the body mass, which we do not know, and scaling of some parameter, characterizing a maximal metabolic capacity, for which we use speed. However, in the discussion, we also consider maneuverability and the need to compensate the force of gravity as other possible parameters, which might add to the maximal metabolic rate. We also do not know the accurate fraction of a maximal metabolic capacity corresponding to the basal metabolic rate.

Speed of birds depends on air resistance and their aerodynamic shapes. The air resistance at the upper range of birds' speed increases faster than linearly, which is a factor also out of our control. The wing movements are well described by the rotational model for limbs that we discussed previously (Eq. (6) in the main text). There are also other important distinguishing features. Mammals operate in two-dimensional, and in linear stretches of chase even in 1-D, environments. Reptiles also act mostly in 2-D space. Fish operates in a 3-D space, but moves mostly horizontally, in a 2-D plane, and fish does not experience force of gravity. Birds inherently operate in all three dimensions of 3-D space, and also experience an additional factor, affecting their metabolism during flight, which is gravitational force they should compensate for using additional metabolic energy. Such are the prerequisites we have for the task.

Data for 30 birds in tabular and graphical forms are presented below.

Table 5D. Maximum birds' mass and wing span, and the maximal speed. Data are from Ref. 38.

| Name | Size, cm | Wing span, *cm* | Mass, *kg* | Speed, $km \cdot h^{-1}$ |
|---|---|---|---|---|
| Quail | 20 | 37 | 0.14 | 24 |
| Guinea_fowl | 71 | 180 | 1.6 | 35 |
| Avocet | 45 | 80 | 0.4 | 40 |
| Barn_Owl | 45 | 110 | 0.55 | 80 |
| Booby | 91 | 155 | 1.8 | 97 |
| Budgerigar | 20 | 35 | 0.04 | 42 |
| Common_Buzzard | 57 | 130 | 1.4 | 40 |
| Duck | 50 | 80 | 1.4 | 55 |
| Goose | 120 | 170 | 8 | 90 |
| Green_Bee-Eater | 18 | 30 | 0.02 | 42 |

| Heron | 140 | 195 | 3 | 64 |
|---|---|---|---|---|
| Keel_Billed_Toucan | 55 | 152 | 4 | 64 |
| Kingfisher | 37.5 | 66 | 0.17 | 40 |
| Long-Eared_Owl | 37 | 98 | 0.3 | 50 |
| Macaw | 100 | 140 | 2 | 24 |
| Magpie | 46 | 60 | 0.25 | 32 |
| Moorhen | 38 | 80 | 0.4 | 35 |
| Nightingale | 16.5 | 22 | 0.22 | 29 |
| Pelican | 106 | 183 | 2.7 | 65 |
| Pheasant | 84 | 86 | 1.5 | 30 |
| Puffin | 32 | 63 | 0.482 | 88 |
| Robin | 14 | 22 | 0.022 | 29 |
| Snowy_Owl | 75 | 164 | 2 | 80 |
| Sparrow | 18 | 20 | 0.042 | 40 |
| Toucan | 63 | 119 | 0.68 | 64 |
| Tawny_Owl | 43 | 105 | 0.65 | 80 |
| Uguisu | 16.5 | 22 | 0.022 | 29 |
| Vulture | 81 | 183 | 2.2 | 48 |
| Woodpecker | 58 | 61 | 0.6 | 24 |

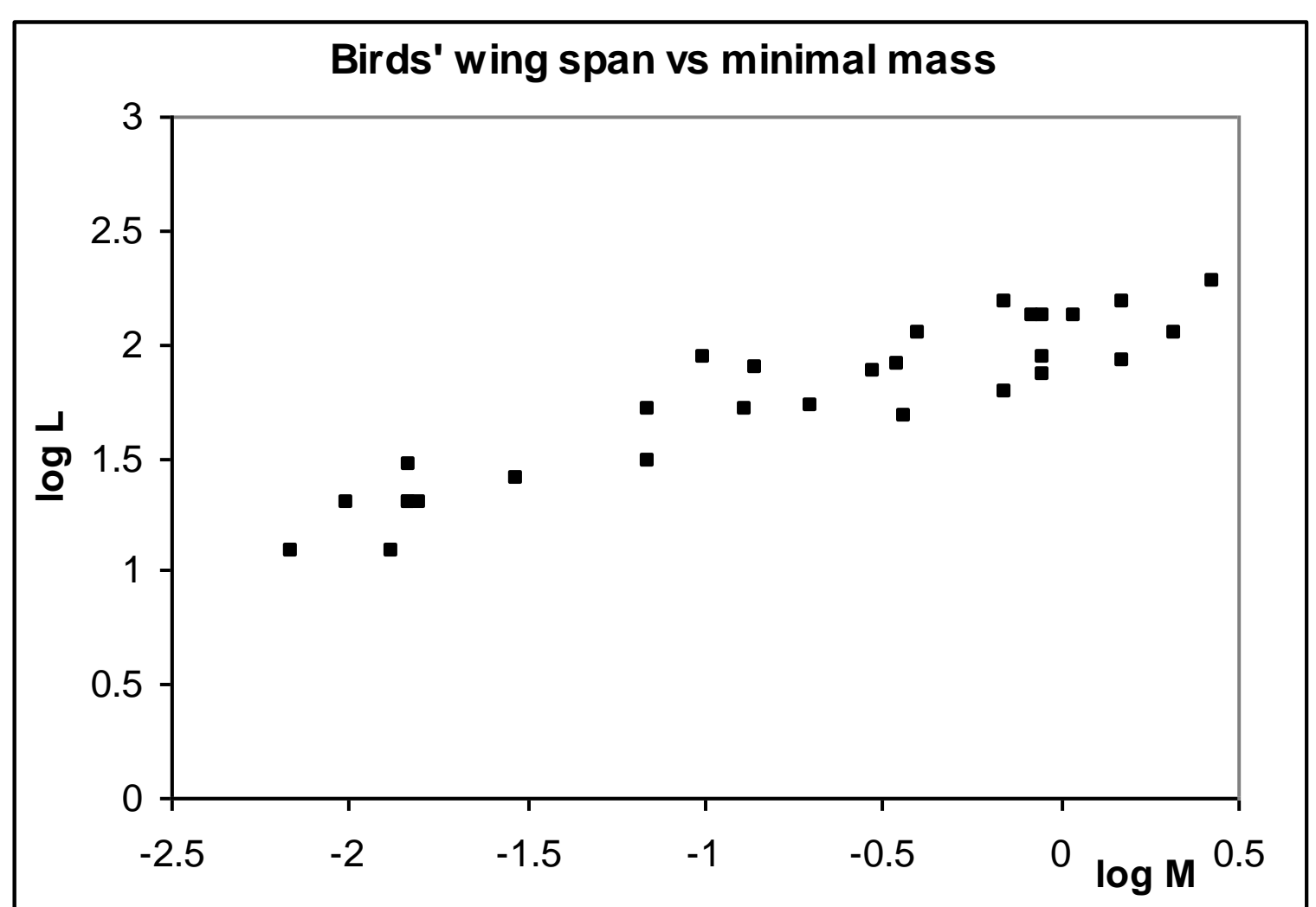

Fig. 5D. Birds' wing span versus minimal mass, in logarithmic scale.

Data for the wing span and speed show well expressed trends and reasonable dispersions. Ref. 38 provides ranges of values for the wing spans and masses, so that we used two separate tables corresponding to maximum and minimum values; thus, effectively, we had 60 entries. Unfortunately, we did not have speeds for smaller birds, so that we used the same speed (given the facts that many birds tend to fly in flocks, and so to maintain the same speed, the difference should not be substantial.) We computed allometric exponents separately for each table (the values differed very little though, 0.396 and 0.402), and then found the average values.

Table 6D. Minimum birds' mass and wing span, and the maximal speed. Data are from Ref. 38.

| Name | Size, cm | Wing span, cm | Mass, kg | Speed, $km \cdot h^{-1}$ |
|---|---|---|---|---|
| Quail | 11 | 30 | 0.07 | 24 |
| Guinea_fowl | 40 | 150 | 0.7 | 35 |
| Avocet | 42 | 77 | 0.14 | 40 |
| Barn_Owl | 25 | 75 | 0.3 | 80 |
| Booby | 64 | 130 | 0.9 | 97 |
| Budgerigar | 15 | 25 | 0.03 | 42 |
| Common_Buzzard | 51 | 110 | 0.4 | 40 |
| Duck | 30 | 60 | 0.7 | 55 |
| Goose | 60 | 83 | 1.5 | 90 |
| Green_Bee-Eater | 16 | 29 | 0.015 | 42 |
| Heron | 85 | 150 | 1.5 | 64 |
| Keel_Billed_Toucan | 42 | 109 | 2.1 | 64 |
| Kingfisher | 10 | 20 | 0.01 | 40 |
| Long-Eared_Owl | 31 | 86 | 0.1 | 50 |
| Macaw | 76 | 86 | 0.9 | 24 |
| Magpie | 40 | 52 | 0.2 | 32 |
| Moorhen | 25 | 50 | 0.07 | 35 |
| Nightingale | 14 | 20 | 0.015 | 29 |
| Pelican | 106 | 183 | 2.7 | 65 |
| Pheasant | 53 | 71 | 0.9 | 30 |
| Puffin | 28 | 47 | 0.368 | 88 |
| Robin | 12.5 | 20 | 0.016 | 29 |
| Snowy_Owl | 60 | 130 | 1.1 | 80 |
| Sparrow | 11.4 | 12 | 0.0134 | 40 |
| Toucan | 29 | 50 | 0.13 | 64 |
| Tawny_Owl | 38 | 81 | 0.35 | 80 |
| Uguisu | 14 | 20 | 0.015 | 29 |
| Vulture | 64 | 130 | 0.85 | 48 |
| Woodpecker | 8 | 12 | 0.007 | 24 |

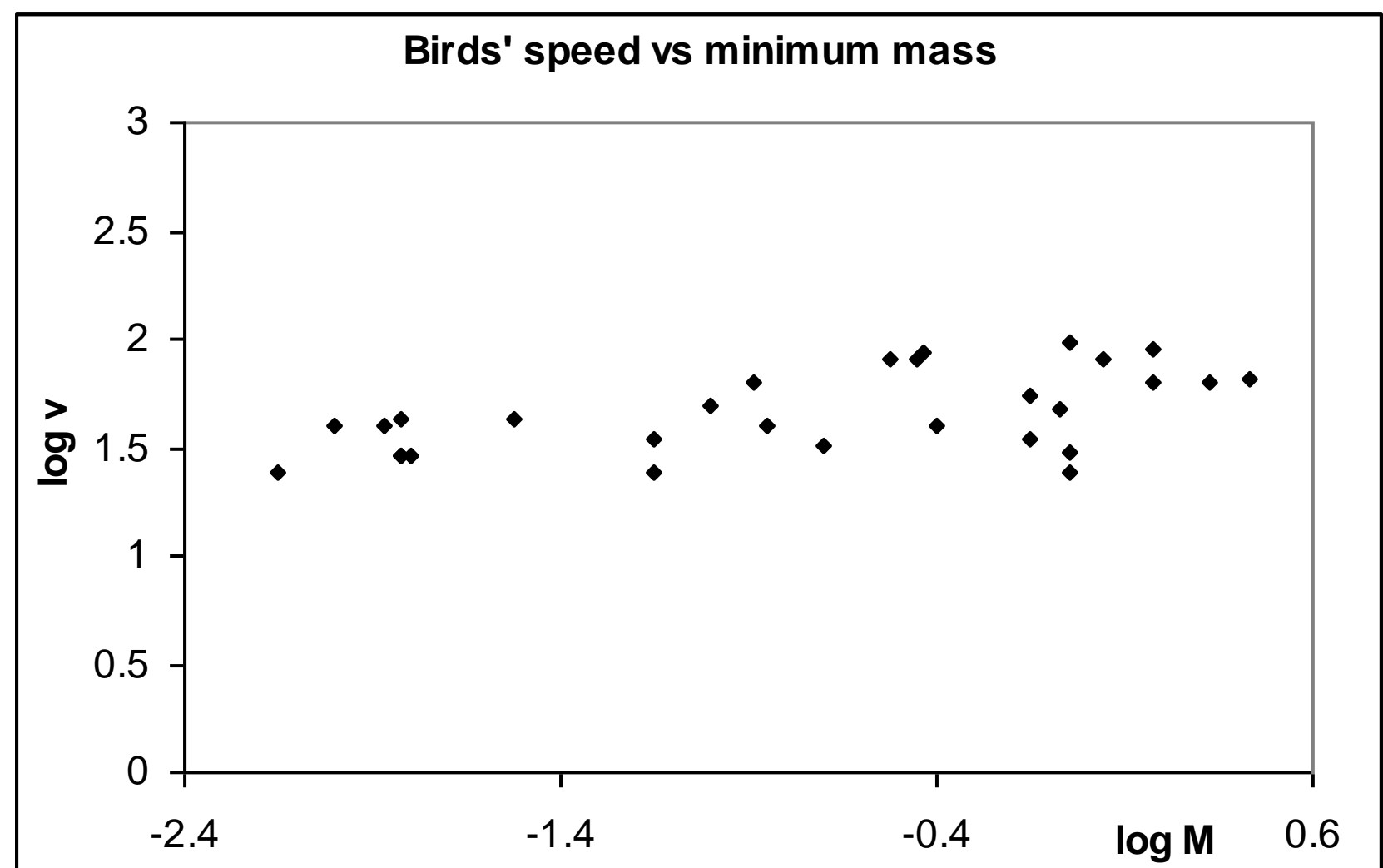

Fig. 6D. Birds' speed versus minimal mass, in logarithmic scale.

The allometric exponent for the wing span relative to mass is $b_l = 0.399 \pm 0.031$. The allometric exponent for the speed relative to mass is $b_v = 0.126 \pm 0.04$. The allometric exponent for the skeletal mass in birds, according to Refs. 1, 51, is in the range of 1.068-1.079. Ref. 52 confirmed but slightly corrected results from Ref. 51 (the authors obtained the value of $1.071 \pm 0.102$ versus the value of $1.079 \pm 0.013$ in Ref. 51). So, we will use a compromise value of 1.073. We assume that the wings' mass scales isometrically with the total mass, similar to mammals. (This is a reasonable assumption, given the considerations presented in the Discussion section. However, we found no concrete studies in this regard.) Then, using Eq. (24), we find

$$b_x = (1 - b_l)b_{sk} + b_v = (1 - 0.399) \times 1.073 + 0.126 = 0.771 \pm 0.05$$

We can estimate the basal metabolic rate similarly to mammals, assuming that $k$=1/10, which gives $b_b = 0.657 \pm 0.041$. Since birds are rather athletic creatures because of the high energetic demand for flight, the value of $k$ can be less. For $k$=1/20 we find $b_b = 0.651 \pm 0.041$. The obtained values agree well with available experimental data for the basal metabolic rate.

Regarding the allometric exponent for the maximal speed, our calculated value of 0.126 matches well the experimental value of 0.13 obtained in Ref. 53 (for a different dataset of birds). Ref. 15 presents the value of 0.644 for the *standard* metabolic rate, which should be slightly lower than the **basal metabolic rate** we calculated, and it is. Still, this number matches our values of 0.651 and 0.657 within the margin of error. Review Ref. 6 quotes results of numerous studies for birds regarding allometric exponents for the basal metabolic rate (0.69, 0.681, 0.667). Ref. 13 gives value of 0.652, and Ref. 42 presents values 0.64 and 0.66, calculated by different statistical approaches, so that our results are right in between. So, for the basal metabolic rate, all these numbers match our values of 0.651 and 0.657 within the margin of error.

Ref. 6 also mentions that a flying hummingbird has an allometric exponent of 0.88-0.95. We did not include it into our data set, since the data point was an outlier located far above the overall trends, both for the wing span and the speed. However, the specific of this point indirectly confirms the high metabolic rate of a hummingbird, whose flight has one special feature - hummingbird can hang in the air, which is supported by specific wing mechanics, and consequently different energy expenditures relative to the total metabolic energy production.

For the maximal metabolic rate Ref. 6 gives values 0.84, 0.88 versus our calculated value of $0.771 \pm 0.05$. The difference is in the range of 0.07-0.11. One of the (rather technical) reasons of this difference is that we excluded high metabolic performers like falcons, eagles from our dataset as outliers with regard to the main array of data, since we did not have sufficient statistics for such birds. (Such non-passerine birds might have lower basal metabolic rate, according to some studies done in 70-s. However, that by no means prevents them to have high maximal metabolic rates). The addition of a statistically representative dataset for such birds will increase the value of $b_v$, without noticeable impact on the scaling of wing span (we verified this assumption by calculations), which will accordingly increase the value of $b_x$. Such considerations are also confirmed by the fact that the value of 0.13 for the allometric exponent for the maximal speed from Ref. 53 is greater than our calculated number 0.126, but only slightly. So, there should be some specific features of flying birds we did not account for, which need energy increase with the growth of birds' mass. This illustrative issue, which also has general importance, is considered in the subsection below.

#### *D.3.1. Factors affecting maximal metabolic rate in birds*

So, we found that in case of birds the presence of other than speed factors is supported by values of obtained allometric exponent, which show noticeable divergence for MMR. Note that it is namely the MMR, where one should expect to find such divergence most visible, since in this case the speed as the main factor contributing to resource acquisition is most relevant. Then, if other factors also noticeably contribute to the value of allometric exponent, using only speed for calculation, as we did, should accordingly produce smaller values of allometric exponent for MMR compared to experimental ones. Indeed, this is what the calculated and experimental values of MMR demonstrate.

On the other hand, the value of the *basal metabolic rate* (BMR), which little relates to speed and other motion characteristics, should not be much affected by incomplete knowledge of how much resources are used to provide energy for the speed in case of birds. Indeed, the obtained results support this consideration.

***Maneuverability***. So, what are the factors we did not account for in calculation of MMR? For now, we can only say that such factors relate to flying at maximal speed. One possible such factor is *maneuverability*. In Ref. 54, the authors say in this regard: "The speed and maneuverability of organisms are central to their fitness, determining the strength and outcome of many species interactions." Combination of speed and maneuverability is an essential "toolset" for birds. These features work together in a way, which enhances birds' survivability. Ref. 55 says: "Larger organisms were faster but less maneuverable than smaller organisms", applying this finding also to birds. In small birds, this is confirmed by decreasing minimal turn radius when mass decreases[55]. Depending on how one looks at this feature and for what purposes, the notion of maneuverability might include not only maneuvers of an entire body, but also of its parts. The last idea is shared by authors of Ref. 55 as well.

Maneuverability, defined in Ref. 55 as a minimum turn radius, follows allometric scaling pattern. The authors acknowledge: "We show the allometry of organismal speed and maneuverability follows predictable scaling principles". However, the minimum turn radius is of little value for our purpose to find *energy* expenditures related to maneuverability.

Maneuvers of birds have curved trajectories. The curvature of trajectories is defined by *centripetal acceleration* (which in turn requires application of centripetal force, and this is where the link to energy appears). During turns, birds use tails and wings for steering. They may still propel themselves during turns using wings, thus directly developing centripetal force, or they may pause wings' motion. When deceleration occurs during turn maneuvers, birds have to *accelerate* to return to the previous speed. This way, both centripetal accelerations and linear accelerations and decelerations relate to maneuverability. Accelerations and decelerations change birds' velocity, and consequently their kinetic energies. (Potential energies change too, when the height of flight changes.) Then, to find the appropriate ratios of energies for birds with different masses, one needs to know how birds' velocities change during turns. However, we do not have such data.

On the other hand, changing direction of motion usually does not require much energy, neither on the ground, nor in the air. As a physical action, changing direction is mostly an elastic one. It changes the *direction* of momentum, but has little influence on its *value*, and consequently energy (think about a ball bouncing from a wall). So, it's doubtful that maneuverability, defined through acceleration, may contribute as much as we need.

***Force of gravity.*** The other unaccounted factor affecting MMR of birds is gravitational force. The action of this force is unique for the birds, since during flight its action should be compensated by application of additional metabolic energy, both in active flight and after gliding (gliding is possible for the cost of speed reduction and decrease of height, unless a bird glides in a raising flow of the air, but this case is irrelevant to our study).

Despite many evolutionary improvements which birds developed during evolution to facilitate compensation of gravitational force, on average the trend that a bigger bird needs to spend more energy for counteracting the force of gravity than a smaller bird is indisputable, and so the allometric exponent for this effect should be positive (assuming the effect is subjected to allometric scaling, which is likely). So, this factor can also increase the total allometric exponent for the MMR of birds. And, given its omnipresence and substantial value of gravitational force, it is likely the one, which could add to our obtained value the amount we need. Detailed study of the role of gravity in metabolism of birds is presented in Ref. 56.

*D.3.2. Propagation of metabolic allometric scaling through the food chain by direct and indirect interaction of species*

There is another important consideration, which has a general importance for understanding principles of formation of metabolic properties of organisms composing food chains, not only for birds.

Many species, birds included, do not feed on each other, but share food resources based on plants, consume small rodents, insects, so that direct speed advantage does not look so critical.

It was explained already that neither the used mathematical approach, nor the main ideas the paper relies upon, consider only pairs of close in mass animals (and birds in particular), related to each other through a food chain. For instance, a wide spectrum of birds-preys have to adjust their metabolic power and escaping skills (in other words, speed and maneuverability) to sustain attacks of birds-predators (which could be the same for many different smaller birds). Accordingly, the combination of thus acquired features is distributed across this wide spectrum of smaller birds - turned out, in accordance with metabolic allometric scaling dependencies. This is one basis, from which allometric scaling originates and propagates through the food chain.

Natural selection in adjusting metabolism, in order to escape predators, is not the only such basis. Supporting feeding capabilities is also important. There is difference in energy expenditures required to catch the same mass of bigger animals (or pick up bigger seeds or nuts) and smaller ones, with the first scenario being energetically more economical. We can show this as follows. Let indexes '1' and '2' relate to a bigger and smaller prey, index '*t*' denotes total value; *E* is energy needed to get the prey; *m* is mass of preys; *b* is an allometric exponent (say, the one accounting for speed difference needed to catch a bigger and smaller prey by the same predator). The total energy required to get the same mass of food by a predator is as follows.

$$E_{1t} = E_1 \, ; \; E_{2t} = E_2 (m_1 / m_2) = E_1 (m_2 / m_1)^b (m_1 / m_2) = E_1 (m_1 / m_2)^{1-b}$$

By initial condition, $m_1 > m_2$. The allometric exponent $b < 1$ (recall, for instance, that for mammals the allometric exponent for velocity was 0.209). Then, we have $E_{2t} > E_{1t}$.

Because of such energy optimization (minimization of energy expenditures), there is a trend to use portions of food preferably contained in a "package" commensurate with the size of a bird (or of another animal), even though other sources of smaller "packages" may not be ignored, depending on circumstances. Such a commensurability of typical portions of food acquired at one time with the size of a bird is another basis, on which metabolic allometric scaling tends to naturally occur and propagate through the food chain. Animal can feed on several sources of food, of course, but the need for more efficient use of energy will move the preference (and accordingly adjusting metabolism) to more commensurate - in portion - size of food sources (which are not rarely substantially more risky to obtain, and nonetheless).

*D.3.3. Concluding birds' study*

We were cautious using speed data for birds. Such, the reader may notice that we did not include into calculations large birds-predators, for several reasons, including the one that some data sources mix diving and flying speeds of birds-predators, like falcons, while the factors driving the diving speed have little relation to metabolism.

We found that this is rather interplay of different mechanisms and factors, affecting distribution of resources between species of a food chain, which create the resulting MAS. Here we considered some additional features affecting MAS of birds, and found that they are also possible factors forming the overall MAS, besides the speed advantage, which is an important, but not the only one such a factor. Moreover, the same consideration, about possible multiplicity of different mechanisms and organismal features forming the total MAS and defining the value of allometric exponent, is entirely applicable to all other classes of animals.

Our purpose was to prove the idea that the optimal sharing of common resources within the food chain, occurring under the guidance of natural selection, is the primary reason and mechanism defining values of allometric exponents. For that, we needed to find classes of animals, which use some dominant feature for resource acquisition, whose characteristics we can measure and quantify in a way, which would allow calculating values of allometric exponents, and compare them with experimental data. Speed was such a parameter; not the only one, but dominant enough in terms of consumed resources to provide calculation of allometric exponents with sufficient accuracy.

## Appendix E. Factors defining interspecific metabolic allometric scaling

### **E.1.** *Scaling of limb masses*

Accounting for this effect is not a trivial issue. We can account for it in a pure form only when limbs move forward in the air. When limbs are in a contact with the ground and provide forward motion of the body, they rotate around the point of contact *R* with the ground, pushing itself and the body forward (Fig. 1E).

The main power goes to keeping the forward motion of the body by application of two forces. One is the reactive force $F_R$, whose horizontal component $F_S$ supports the forward motion, while the vertical component $F_W$ counterbalances the weight (being greater when the body moves up, and less when down). The force $F_A$ is needed to lift the body slightly up at the beginning of forward movement, rotating it around the point *B* counterclockwise, and then to support the body preventing it from falling until the front legs will provide support for the body.

Thus, we can view this movement as synchronized rotations of limbs around point *R* and the body around point *B*. In this decomposition of forces, it is the body, which requires horizontal acceleration and counterbalancing of weight, that will eventually consume the most of energy, while the limbs, although being providers of large part of required power, need less energy to accelerate and lift themselves (as it was mentioned before, from 8 to 33% of the total locomotor costs).

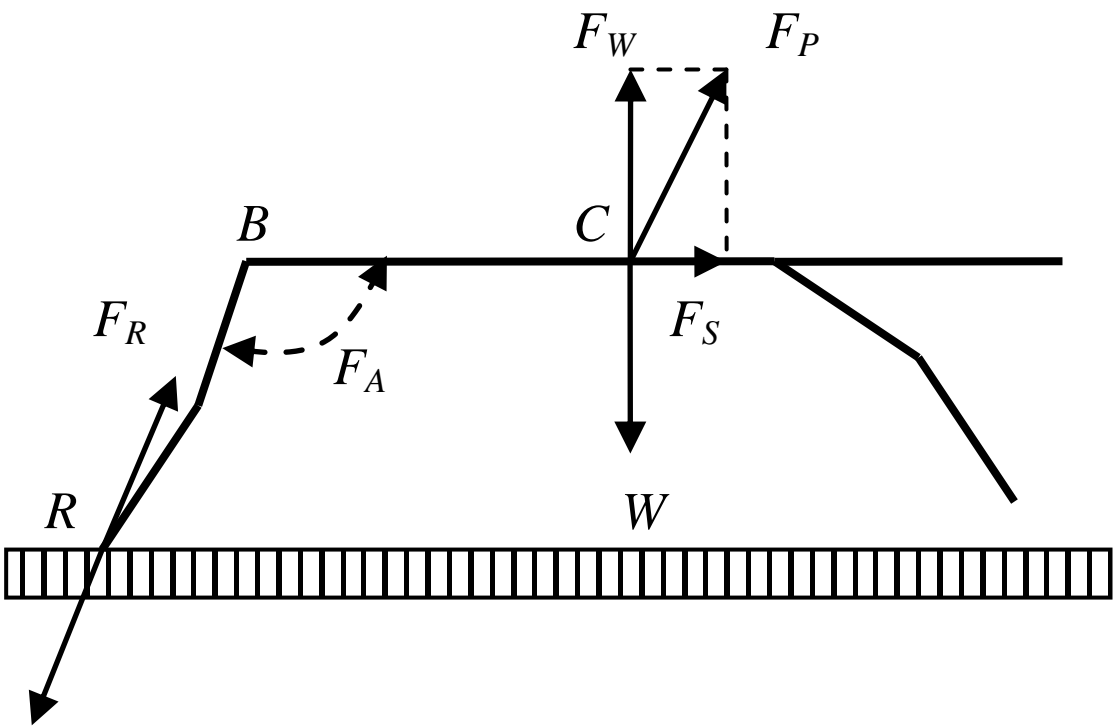

Fig. 1E. Mechanical forces acting on the limbs and body during animal movement. *W* - weight; $F_W$ - the vertical component of a reactive force counteracting the body weight *W* at a center of mass *C*; $F_S$ - horizontal component of a reactive force providing horizontal acceleration; force $F_A$ pushes apart the body and the limbs rotating the body around point *B* relative to limbs; $F_R$ is a reactive force counteracting the force of pushing the ground.

In this arrangement, from the point of view of inertial forces, it is the length of limbs, to the upper part of which the body is attached, is of greater importance for the moment of inertia, than the limb mass. Of course, it would be useful to know the exact scaling of limb mass, like in case of mammals, but if it is unknown, given the above considerations, the assumption about the proportional increase of limb masses to the total mass (called geometric similarity in the literature) will not introduce substantial error. Accounting for the distribution of energy between the movements of body and limbs can be done with reasonable accuracy even on the basis of the model presented in Fig. 1E, if needed, from which the scaling of this component can be accessed. If we denote such an allometric exponent as $b_m$, then Eq. (24) transforms to the following.

$$b_x = (b_m - b_l) b_{sk} + b_v$$

However, as it was said above, for many practical purposes one can assume that $b_m = 1$.

### E.2. *Distribution of inertial masses*

Distribution of inertial masses between moving limbs means not only masses of limbs, but also distribution of the body mass relative to acceleration thrusts. Forelimbs provide more support and cushion for the landing, while the hindlimbs are the main propellers in animals[32]. For our purposes, that fact was not important, since we used an average length of limbs. However, if the difference in fore- and hindlimbs is substantial, or high accuracy is required, then the specifics of mass distribution should be taken into account.

**E.3.** ***Scaling of limb lengths***

Scaling of limb lengths (tail in fish), as we have seen for all considered animals, is an important parameter, significantly affecting the value of allometric exponents. For the typical proportions in animals, this parameter is far more important than scaling of limb mass. Geometry of limb movement (in particular, the striding angles) is also a factor that can noticeably affect the overall energy expenditures. Fortunately, many animals use similar geometrical patterns for movement, which in many instances was evolutionarily optimized by similar environmental factors, like force of gravity, hydraulic resistance, landscape, terrain, etc. The similarity of striding angles for mammals was confirmed by studies in Ref. 32.

**E.4.** ***Scaling of skeleton mass***

Scaling of skeleton mass, as our study showed, turned out to be an important factor for all considered taxa. We saw how such an uncertainty in case of reptiles affected the value of allometric exponents. Fortunately, in some instances it can be found through other parameters, which we did using Eq. (24). Scaling of skeleton mass occurs differently in different species, although mechanical constraints seem as one of the main reasons of this scaling.

**E.5.** ***Scaling of the maximal metabolic power***

Scaling of the maximal metabolic power is probably one of the most important and difficult to find parameters. The reason is that the metabolic power is manifested in many different ways. We saw that in small lizards the actual metabolic power was 3.9 times greater than one could deduce from the running speed[39].

As we saw in case of mammals and fish, the maximal speed is an adequate measure of metabolic power only when this is the major factor supporting animal existence. In case of some animals, like elephants, the speed is not necessarily of such high importance, and so in such cases speed as a measure of the maximal metabolic power is not an accurate parameter. However, in general, the maximal speed is a good indication of the maximal metabolic power, as we have seen in our study. The thing is how to get animals to run, or fish to swim, at a maximum speed. Overall, the area of maximal metabolic output and the ways of its manifestation is a very promising, interesting and practical one, which awaits studies.

**E.6.** ***Finding fractions of the maximal metabolic power***

Finding fractions of the maximal metabolic power corresponding to basal and other specific metabolic rates is an important problem. It can provide many scientific and practical insights into organismal physiology and adaptation means for different organisms living in different environments, as well as for evaluation of other metabolic characteristics, examples of which we presented in our study.